\documentclass[12pt]{article}

\usepackage{jheppub}
\usepackage{amsmath, amssymb, amsfonts,lipsum}
\usepackage{hyperref}
\usepackage{graphicx, subfigure, epstopdf}
\usepackage{color}

\newcommand{\be}{\begin{equation}}
\newcommand{\ee}{\end{equation}}

\title{The Higgs Vacuum Uplifted: \\ Revisiting the Electroweak Phase Transition with a Second Higgs Doublet}
\author[a]{G.~C.~Dorsch,}
\author[b]{S.~J.~Huber,}
\author[b,c]{K.~Mimasu,}
\author[b,d]{J.~M.~No}
\affiliation[a]{DESY, Notkestra\ss e 85, D-22607 Hamburg, Germany}
\affiliation[b]{Department of Physics and Astronomy, University of Sussex, Brighton, BN1 9QH, UK }
\affiliation[c]{Center for Cosmology, Particle Physics and Phenomenology - CP3, Universit\'e Catholique de Louvain, Louvain-la-neuve, Belgium }
\affiliation[d]{Department of Physics, King's College London, Strand, WC2R 2LS London, UK}

\emailAdd{glauber.carvalho.dorsch@desy.de}
\emailAdd{S.Huber@sussex.ac.uk}
\emailAdd{ken.mimasu@uclouvain.be}
\emailAdd{jose$\_$miguel.no@kcl.ac.uk}
 
\abstract{The existence of a second Higgs doublet in Nature could lead to a cosmological first order electroweak 
phase transition and explain the origin of the matter-antimatter asymmetry in the Universe. We explore the parameter space 
of such a two-Higgs-doublet-model and show that a first order electroweak phase transition strongly correlates with a significant uplifting of the Higgs vacuum w.r.t. 
its Standard Model value. We then obtain the spectrum and properties 
of the new scalars $H_0$, $A_0$ and $H^{\pm}$ that signal such a phase transition, showing that the decay $A_0 \rightarrow H_0 Z$ at the 
LHC and a sizable deviation in the Higgs self-coupling $\lambda_{hhh}$ from its SM value are
sensitive indicators of a strongly first order electroweak phase transition in the 2HDM.}

\begin{document}

\titlepage

\maketitle

\newpage

\section{Introduction}

In a cold, nearly empty Universe, spontaneous breaking of the electroweak (EW) symmetry takes place because the Higgs potential energy 
is minimized when the Higgs field(s) acquire non-vanishing vacuum expectation values (VEVs). But in the early Universe, when the 
scalar fields are surrounded by a thermal plasma of particles, 
the net free-energy of the entire system has further contributions stemming from interactions with this thermal bath, which yield 
a restoration of the EW symmetry for temperatures $T \gtrsim 100$ GeV. 
Tracing the thermal history of the Higgs field from the high temperature regime  down to the $T=0$ vacuum of today reveals the properties of the Electroweak Phase Transition (EWPT), the process of EW 
symmetry breaking in the early Universe.

The detailed dynamics of the EWPT is a crucial ingredient for a number of cosmological observables. One example 
is the baryon asymmetry of the Universe (BAU), which could be dynamically generated during a first order EWPT as long as the nucleation and expansion 
of vacuum bubbles provide a strong enough departure from thermal equilibrium as required by the Sakharov conditions~\cite{Sakharov:1967dj}.
For the minimal Higgs sector of the SM, a first order transition would only be achieved for a Higgs mass $m_h$ lower than the mass of the $W$ boson, 
$m_h\lesssim m_W$~\cite{Kajantie:1996mn}, and thus does not occur in the SM~\cite{D'Onofrio:2014rug}. 
The BAU therefore constitutes concrete evidence of physics beyond the SM which can be connected to the EWPT and the precise 
nature of the Higgs sector. In addition, a first order EWPT would generate a stochastic background of gravitational waves, 
potentially observable with the upcoming space-based gravitational wave interferometer LISA (see~\cite{Caprini:2015zlo} for a review). 
Since the properties of the EWPT are highly sensitive to the presence of new degrees of freedom at the EW scale coupling to the Higgs field, its study provides a tantalising research topic at the interface of particle physics and cosmology, 
shedding light on the ultimate structure of the sector responsible for EW symmetry breaking in Nature. This is 
a key goal of the physics programme of the LHC and future colliders.

To fully determine the nature of the EWPT one typically has to inspect the shape and evolution of the Higgs thermal effective potential with temperature, 
which faces various theoretical issues (see e.g.~\cite{Patel:2011th,Garny:2012cg,Curtin:2016urg}). 
Furthermore, determining the phase transition strength is usually a computationally expensive algorithm.
On the other hand, it has been recently pointed out that, in theories where a modified 
scalar sector acts as the main source of a strong phase transition, the EWPT strength is 
closely correlated with the zero temperature vacuum energy difference of the theory~\cite{Huang:2014ifa,Harman:2015gif}. 
The amount by which the EW broken vacuum is ``uplifted'' with respect to the SM case 
constitutes a good indicator of the increase in the strength of the EWPT. 

In this work we will investigate this correlation in the context of two-Higgs-doublet models (2HDMs)~(see~\cite{Branco:2011iw} for a review). 
Despite the minimality of the model, the existence of additional scalars can induce a strongly first order phase transition~\cite{Cline:1996mga, Dorsch:2013wja, Dorsch:2014qja, Basler:2016obg}, as well as 
introduce new sources of Charge-Parity (CP) violation to enable the successful generation of the BAU via EW baryogenesis in some regions of its 
parameter space~\cite{Fromme:2006cm, Cline:2011mm, Dorsch:2016nrg}. 
Ultimately, lattice calculations will provide a detailed map of the 2HDM parameter region in which a strong first order EWPT occurs, 
but perturbative calculations can already point to the main features of such a map.
We show that the correlation between the EWPT strength and the zero temperature vacuum energy uplifting
is a powerful analytic tool to explore the interplay between experimental/theoretical constraints 
and the strength of the EWPT in 2HDM scenarios.

Our analysis indicates that this interplay results in a strong EWPT favouring 
a hierarchical 2HDM scalar spectrum, with a preference for a heavy 
charged and pseudoscalar as compared to the neutral scalars (which includes the 125 GeV Higgs boson). 
This leads to a ``smoking-gun'' signature at the LHC~\cite{Dorsch:2014qja} (see also~\cite{Coleppa:2014hxa,Kling:2016opi}). 
We also show a significant deviation of the Higgs self-coupling from its SM 
value to be a collateral prediction of 2HDM scenarios with a strong EWPT~\cite{Kanemura:2002vm, Kanemura:2004ch}. Accessing the Higgs 
self-coupling is a key goal of the LHC and future colliders (see e.g.~\cite{Dolan:2012rv,Barr:2013tda,McCullough:2013rea,Barr:2014sga} 
for recent analyses), as it provides a direct probe of the nature of EW symmetry breaking. In the High-Luminosity LHC the sensitivity of such measurement 
is expected to be $\sim 50$~\%~\cite{Yao:2013ika, Dawson:2013bba}. 
We will show that this could be enough to probe some scenarios with a strong EWPT in 2HDMs.

In section~\ref{sec:2HDM_Review} we provide a review of the 2HDM and establish our notation as well as the relevant theoretical constraints on the 
model parameters. Section~\ref{sec:EWPT} elaborates on the computation of the vacuum energy difference 
in the 2HDM. Section~\ref{sec:scan2} presents the numerical scan of the 2HDM parameter space, establishing the correlation between the 
vacuum energy difference and the strength of the EWPT,  
well as highlighting a number of key features of 2HDMs that exhibit strongly first order EWPTs. We move to a more analytical treatment in 
section~\ref{sec:scan}, using the vacuum energy difference as a proxy for the phase transition strength, delving deeper into the effects that establish 
the preferred regions of parameter space. Section~\ref{sec:trilinear} discusses the connection of the phase transition with the trilinear Higgs 
self-coupling before conclusions are drawn in section~\ref{sec:conclusions}.

\section{Reviewing Two Higgs Doublet Scenarios}
\label{sec:2HDM_Review}

Let us start with a brief review of the 2HDM, which also defines our notation in this work. We consider a 2HDM scalar potential with 
a softly broken $\mathbb{Z}_2$ symmetry to inhibit tree-level flavour changing neutral currents (FCNC), and for simplicity we neglect effects 
from CP violation\footnote{CP violation is important for the computation of the final baryon asymmetry, but its impact on the phase transition 
strength is typically negligible, as EDM constraints require the CP violating phase to be small~\cite{Dorsch:2016nrg,Inoue:2014nva}.}. The scalar 
potential then reads  
\begin{eqnarray}	
\label{2HDM_potential} 
V_{\mathrm{tree}}(\Phi_1,\Phi_2) &= &\mu^2_1 \left|\Phi_1\right|^2 + \mu^2_2\left|\Phi_2\right|^2 - \mu^2\left[\Phi_1^{\dagger}\Phi_2+\mathrm{h.c.}\right] +
\frac{\lambda_1}{2}\left|\Phi_1\right|^4 +\frac{\lambda_2}{2}\left|\Phi_2\right|^4   \nonumber \\
&+& \lambda_3 \left|\Phi_1\right|^2\left|\Phi_2\right|^2 + \lambda_4 \left|\Phi_1^{\dagger}\Phi_2\right|^2 + 
\frac{\lambda_5}{2}\left[\left(\Phi_1^{\dagger}\Phi_2\right)^2+\mathrm{h.c.}\right]\, ,
\end{eqnarray}
where the two scalar $SU(2)_L$ doublets $\Phi_j$ ($j = 1,2$) may be written as 
\begin{equation}
\label{doublets}
\Phi_k = \left(\begin{array}{c}
\phi_k^{+} \\
\frac{v_k + \varphi_k + i\,\eta_k}{\sqrt{2}}
\end{array}
\right).
\end{equation}
The physical scalar sector of a 2HDM is comprised of two CP-even neutral scalars, $h$ and $H_0$ (with $m_{H_0} \geq m_h$), 
plus a neutral CP-odd scalar $A_0$ and a charged scalar $H^{\pm}$. 
In this work we identify $h$ with the observed 125 GeV Higgs boson, but we stress that our main arguments can be easily extended
to the flipped case where $H_0$ is the recently observed particle and $h$ is a lighter and yet undetected scalar (experimental constraints on 
this scenario have been recently discussed in~\cite{Bernon:2015wef,Brooijmans:2016vro,Aggleton:2016tdd}).

Apart from $m_h$ and $v = 246$ GeV, the scalar potential~\eqref{2HDM_potential} may be parametrized in terms of $\mathrm{tan}\,\beta \equiv v_2/v_1$
(with $v^2_1 + v^2_2 = v^2$), the angle $\alpha$ parametrising the mixing between the CP-even states,
the scalar masses $m_{H_0}$, $m_{A_0}$, $m_{H^{\pm}}$ and the mass scale $M$,
\begin{equation}
	M^2 \equiv \mu^2(t_\beta+t_\beta^{-1}).
\end{equation}
The relation between the physical states $h, \,H_0, \,A_0, \,H^{\pm}$ and 
the states $\varphi_j,\, \eta_j,\, \phi_j^{\pm}$ is given by 
\begin{eqnarray}
\label{rotation_states}
H^{\pm}=-s_{\beta}\, \phi_1^{\pm} + c_{\beta}\, \phi_2^{\pm}, & \hspace{9mm} & A_0=-s_{\beta}\, \eta_1 + c_{\beta}\, \eta_2, \nonumber \\
h=-s_{\alpha}\, \varphi_1 + c_{\alpha}\, \varphi_2,  & \hspace{9mm} & H_0=-c_{\alpha}\, \varphi_1 - s_{\alpha}\, \varphi_2, \nonumber
\end{eqnarray}
with $s_{\beta},c_{\beta}, s_{\alpha},c_{\alpha} \equiv \mathrm{sin}\,\beta, \mathrm{cos}\,\beta, \mathrm{sin}\,\alpha, \mathrm{cos}\,\alpha$, respectively. 
Regarding the couplings of the two doublets $\Phi_{1,2}$ to fermions, the $\mathbb{Z}_2$ symmetry in (\ref{2HDM_potential}), even when softly 
broken by $\mu^2$, may be used to
forbid potentially dangerous tree-level FCNCs by requiring that each fermion type couple to 
one doublet only~\cite{Glashow:1976nt}. 
By convention, up-type quarks couple to $\Phi_{2}$. In Type I 2HDM all the other fermions 
also couple to $\Phi_{2}$, while for Type II down-type quarks and leptons couple to $H_{1}$. There are two more possibilities (depending on the 
$\mathbb{Z}_2$ parity assignment for leptons with respect to down-type quarks), but we focus here on Types I and II. 
The parameters $t_{\beta} \equiv \mathrm{tan}\,\beta$ and $c_{\beta -\alpha} \equiv \mathrm{cos}\,(\beta-\alpha)$ control the strength of the couplings 
of $h$, $H_0$, $A_0$ and $H^{\pm}$ to gauge bosons and fermions. In particular, one can identify the so-called alignment limit~\cite{Gunion:2002zf} 
$c_{\beta-\alpha} = 0$, for which $h$ couples to SM particles exactly like the SM Higgs. The parameters in the scalar potential can be related to the masses 
and mixings in the scalar sector as shown in Appendix~\ref{app:dictionary}. 

In order to obtain a viable 2HDM scenario, theoretical constraints from unitarity, perturbativity and stability/boundedness 
from below of the scalar potential (\ref{2HDM_potential}) need to be satisfied. These will play an important role in the following discussion. 
Tree-level boundedness from below of the potential~\eqref{2HDM_potential} requires
\begin{equation}
 \lambda_1>0\,, \quad \,\, \lambda_2>0\,,  \quad \,\,
\lambda_3>-\sqrt{\lambda_1 \lambda_2}\,, \quad\,\, \lambda_3+\lambda_4 - |\lambda_5|>-\sqrt{\lambda_1 \lambda_2}\,.
 \label{eq:vacstability}
\end{equation}
At the same time, tree-level unitarity\footnote{For a recent one-loop analysis, leading to slightly more stringent bounds, see~\cite{Grinstein:2015rtl}.}
imposes bounds on the size of various combinations of the quartic couplings $\lambda_i$~\cite{Akeroyd:2000wc,Ginzburg:2005dt}.
Similar (although generically less stringent) bounds on $\lambda_i$ may be obtained from perturbativity arguments.
Finally, in order to guarantee absolute tree-level stability of the EW minimum (that is, 
the non-existence of a ``panic vacuum"~\cite{Barroso:2013awa, Ivanov:2015nea}), the couplings must satisfy
\begin{equation}
	\left[ \left( \frac{m^2_{H^\pm}}{v^2} + \frac{\lambda_4}{2}\right) - \frac{|\lambda_5|^2}{4}\right]
	\left[ \frac{m^2_{H^\pm}}{v^2}  + \frac{\sqrt{\lambda_1\, \lambda_2}-\lambda_3}{2}\right] > 0,
\end{equation}
which can be rewritten as
\begin{equation}
\label{panicvacuum}
	\small
	\frac{M^2 m_{A_0}^2}{2v^4}
	\left\{\frac{M^2}{v^2} + \frac{(m_{H_0}^2 - m_h^2)}{v^2}\left[ s_{\beta-\alpha}^2 - c_{\beta-\alpha}^2 - c_{\beta-\alpha}\,s_{\beta-\alpha} 
	(t_\beta - t_\beta^{-1})\right] + \sqrt{\lambda_1\, \lambda_2} \right\} > 0.
\end{equation}
Note that, in alignment, the condition that no panic-vacua exist at tree-level is satisfied for $M^2 > 0$.

\vspace{2mm}

In the following, it will prove convenient to use the Higgs basis of the 2HDM~\cite{Gunion:2002zf}, given by the rotation from 
the doublet fields in~\eqref{doublets} via
\begin{eqnarray}
H_1 & = &    c_{\beta}\, \Phi_1 + s_{\beta}\, \Phi_2,  \nonumber \\
H_2 & = &  - s_{\beta}\, \Phi_1 + c_{\beta}\, \Phi_2 \,.
\end{eqnarray}
The two doublets in the Higgs basis read
\begin{equation}
\label{doublets_mass}
H_1 = \left(\begin{array}{c}
G^{+} \\
\frac{v + h_1 + i\,G_0}{\sqrt{2}}
\end{array}
\right)
\quad \quad , \quad \quad
H_2 = \left(\begin{array}{c}
H^{+} \\
\frac{h_2 + i\,A_0}{\sqrt{2}}
\end{array}
\right),
\end{equation}
such that the EW broken phase is characterized by $\langle h_1 \rangle = v$, $\langle h_2 \rangle = 0$, with 
$h_1$, $h_2$ the CP even field directions of $H_1$ and $H_2$.
The 2HDM tree-level potential for $H_i$ reads
\begin{eqnarray}	
\label{2HDM_potential2}
V_{\rm tree}(H_1,H_2)= \bar{\mu}^2_1 \left|H_1\right|^2 + \bar{\mu}^2_2\left|H_2\right|^2 - \bar{\mu}^2 
\left[H_1^{\dagger}H_2+\mathrm{H.c.}\right] +\frac{\bar{\lambda}_1}{2}\left|H_1\right|^4 \nonumber \\
 +\frac{\bar{\lambda}_2}{2}\left|H_2\right|^4 + \bar{\lambda}_3 \left|H_1\right|^2\left|H_2\right|^2
+\bar{\lambda}_4 \left|H_1^{\dagger}H_2\right|^2+
 \frac{\bar{\lambda}_5}{2}\left[\left(H_1^{\dagger}H_2\right)^2+\mathrm{H.c.}\right] \nonumber \\
 + \bar{\lambda}_6 \left[  \left|H_1\right|^2 H_1^{\dagger}H_2 +\mathrm{H.c.}\right] + \bar{\lambda}_7 \left[ 
 \left|H_2\right|^2 H_1^{\dagger}H_2 +\mathrm{H.c.}\right]\,,
\end{eqnarray}
with the modified mass parameters $\bar{\mu}^2_1$, $\bar{\mu}^2_2$, $\bar{\mu}^2$ and quartic couplings $\bar{\lambda}_{1-7}$ being functions of 
$m^2_{H^\pm}$, $m^2_{A_0}$, $m^2_{H_0}$, $m^2_{h}$, $M^2$, $c_{\beta-\alpha}$ and $t_{\beta}$ (see Appendix~\ref{app:HBdict}). We also note that in the Higgs basis 
$M$ precisely corresponds to the mass scale of the second doublet prior to EW symmetry breaking.

\section{The Electroweak Phase Transition with Two Higgs Doublets \label{sec:EWPT}}

\vspace{-2mm}

The evolution of the Higgs vacuum in the early Universe, in thermal equilibrium, 
can be described by means of the finite temperature effective potential $V^{T}_{\mathrm{eff}}(\phi,T)$ for the Higgs 
(and possibly other scalar fields subject to evolution in the early Universe)
\be
V^{T}_{\mathrm{eff}}(\phi,T) = V_{\rm tree}(\phi) + V_1(\phi) + V_T(\phi,T) \, ,
\ee
with $\phi$ representing the set of relevant scalar fields including the Higgs, $V_{1}$ being the $T=0$ radiative Coleman-Weinberg piece of the
effective potential and $V_T$ the thermal contribution. The free-energy density difference $\mathcal{F}_T$ between the 
$SU(2)_{\mathrm{L}} \times U(1)_{\mathrm{Y}}$ symmetric phase $\left\langle \phi \right\rangle = 0$ and the 
broken phase $\left\langle \phi \right\rangle = v_T \neq 0$ at temperature $T$ is then
\be\begin{split}
\label{F_T}
\mathcal{F}_T =&~V_{\rm eff}^T(v_T,T) - V_{\rm eff}^T(0,T) \\
			\equiv &~ \mathcal{F}_0 + V_0(v_T) - V_0(v_0) + V_T(v_T,T) - V_T(0,T) = \mathcal{F}_0 + \Delta V_T\, .
\end{split}\ee
The first contribution, $\mathcal{F}_0<0$, corresponds to the vacuum energy difference at $T=0$, while the 
second contribution $\Delta V_T \geq 0$ is monotonically increasing with $T$, vanishing as $T$ vanishes. 
The critical temperature, $T_c$, below which the EWPT can proceed in the early Universe is then defined by $\mathcal{F}_{T_c} = 0$.

A first order EWPT is characterized by the presence of a potential barrier between the symmetric and broken phases as 
$\mathcal{F}_T$ turns negative during the evolution of the Universe. Such a first order transition 
could be responsible for the generation of the matter-antimatter asymmetry of the Universe through EW baryogenesis, should the strength of the 
transition be sufficiently large (see~\cite{Quiros:1994dr,Morrissey:2012db,Konstandin:2013caa} for reviews on the EWPT and baryogenesis).
The details of the tunneling process~\cite{Coleman:1977py,Linde:1980tt,Linde:1981zj} between 
symmetric and broken phases in a first order EWPT depend on the functional form of $\Delta V_T$ 
in~\eqref{F_T}. 
Nevertheless, it has been recently shown that in a wide class of extensions of the SM potentially leading to a first order EWPT, 
the strength of the transition, which is the relevant quantity for EW baryogenesis, is dominantly controlled by the value of 
$\mathcal{F}_0$ {\it w.r.t.}~its corresponding value for the SM, $\mathcal{F}^{\mathrm{SM}}_0$~\cite{Huang:2014ifa,Harman:2015gif}. 
In this work we show that this is indeed the case for the 2HDM. It is then possible to perform a systematic study 
of the 2HDM parameter space in which a strongly first order EWPT is favoured by analyzing the behaviour of 
$\Delta \mathcal{F}_0 \equiv \mathcal{F}_0 - \mathcal{F}^{\mathrm{SM}}_0$. Moreover, we stress that 
$\Delta \mathcal{F}_0$ is renormalization scale independent and safe from potential gauge dependence issues~\cite{Patel:2011th,Garny:2012cg}, 
being manifestly gauge invariant. These highlight the advantage of using $\Delta \mathcal{F}_0$ to explore 
the regions of 2HDM parameter space where a strongly first order EWPT is possible, as well as its phenomenological implications.

Let us now discuss the vacuum energy at 1-loop in 2HDM scenarios. 
For the renormalization of the 2HDM 1-loop effective potential we use an on-shell scheme, imposing (among other conditions) that the
value of the 1-loop vevs for the two doublets and the 1-loop physical masses $m_h$, $m_{H_0}$, $m_{A_0}$ and $m_{H^{\pm}}$ 
are equal to their tree-level values. The renormalized 1-loop effective potential in the Higgs basis reads
\begin{eqnarray}	
\label{2HDM_potential3}
V_{\rm tree}(H_1,H_2) + V_{\rm CT}(H_1,H_2) + V_{\rm 1},
\end{eqnarray}
with the counterterm potential being
\begin{eqnarray}	
\label{2HDM_potential4}
V_{\rm CT}(H_1,H_2) = -\delta\bar{\mu}^2_1 \left|H_1\right|^2 + \delta\bar{\mu}^2_2\left|H_2\right|^2 - \delta\bar{\mu}^2 
\left[H_1^{\dagger}H_2+\mathrm{H.c.}\right] +\frac{\delta\bar{\lambda}_1}{2}\left|H_1\right|^4 \nonumber \\
 +\frac{\delta\bar{\lambda}_2}{2}\left|H_2\right|^4 + \delta\bar{\lambda}_3 \left|H_1\right|^2\left|H_2\right|^2
+\delta\bar{\lambda}_4 \left|H_1^{\dagger}H_2\right|^2+
 \frac{\delta\bar{\lambda}_5}{2}\left[\left(H_1^{\dagger}H_2\right)^2+\mathrm{H.c.}\right] \nonumber \\
 + \delta\bar{\lambda}_6 \left[  \left|H_1\right|^2 H_1^{\dagger}H_2 +\mathrm{H.c.}\right] + \delta\bar{\lambda}_7 \left[ 
 \left|H_2\right|^2 H_1^{\dagger}H_2 +\mathrm{H.c.}\right]\,.
\end{eqnarray}
An immediate advantage of working in the Higgs basis is that, in order to
obtain the vacuum energy $\mathcal{F}_0$, we only need to compute the on-shell renormalization conditions 
explicitly\footnote{The Higgs basis condition $\langle h_2 \rangle = 0$ is maintained at 1-loop by the choice of $\delta \bar{\mu}^2$ 
and $\delta \bar{\lambda}_6$.}
for $\delta \bar{\mu}_1^2$ and $\delta \bar{\lambda}_1$
\begin{eqnarray}	
\label{2HDM_counterterms}
-\delta \bar{\mu}_1^2 + \frac{\delta \bar{\lambda}_1 \,v^2}{2} + \frac{1}{v} \left. \frac{\partial V_1}{\partial h_1} \right|_v =  0 \quad\,\,,\,\, \quad 
-\delta \bar{\mu}_1^2 + \frac{3\,\delta \bar{\lambda}_1 \,v^2}{2} + \left. \frac{\partial^2 V_1}{\partial h_1^2} \right|_v =  0\,.
\end{eqnarray}
The 1-loop piece of the scalar potential $V_1$ in~\eqref{2HDM_potential3} is given in Landau gauge (see {\it e.g.}~\cite{Cline:2011mm}) by
\begin{eqnarray}	
\label{2HDM_potential5}
V_{\rm 1} = \sum_{\alpha} n_{\alpha} \frac{m_{\alpha}^4(h_1,h_2)}{64\pi^2}\left(\log\frac{|m_{\alpha}^2(h_1,h_2)|}{Q^2} - C_{\alpha}\right)\, .
\end{eqnarray}
The index $\alpha$ sums over $W,\,Z$ gauge bosons, top quark and 2HDM scalars including Goldstone 
bosons\footnote{We note the squared masses of the scalars do \emph{not} vanish at the origin in general. As these masses may be negative for certain 
values of $h_1$, $h_2$, the absolute value in the argument of the logarithm ensures only the real part of the potential is evaluated.}, 
with $n_{\alpha} > 0$ ($n_{\alpha} < 0$) for bosons (fermions). The various $C_\alpha$ are constants which depend on the renormalization scheme, 
and may be disregarded as they drop out in the following analysis. 
The vacuum energy $\mathcal{F}_0$ reads 
\begin{eqnarray}	
\label{F_0}
\mathcal{F}_0 = -\frac{m_h^2 v^2}{8} - \frac{v^2}{8}c^2_{\beta-\alpha}\,(m_{H_0}^2 - m_h^2) + \Delta V_1 - \frac{\delta \bar{\mu}_1^2 \,v^2}{2} 
+ \frac{\delta \bar{\lambda}_1 \,v^4}{8},
\end{eqnarray}
where $\Delta V_1$ is to be understood as the difference of the Coleman-Weinberg terms~\eqref{2HDM_potential5} evaluated at the 
electroweak minimum and at the origin. 
As we are ultimately interested in $\Delta \mathcal{F}_0$, 
we also need to compute $\mathcal{F}^{\mathrm{SM}}_0$ using the same on-shell renormalization procedure
(demanding the 1-loop Higgs vev and mass to match their tree level values), obtaining
\be 
\label{F_0_SM}
\mathcal{F}_0^{\rm SM} = -\frac{m_h^2 v^2}{8}  + \frac{1}{64\pi^2}\left(3m_W^4+ \frac{3}{2} m_Z^4-6 m_t^4 \right) + \frac{m_h^4}{64\pi^2}\left( 3 + \log 2\right)\,.
\ee
The first term in~\eqref{F_0} and~\eqref{F_0_SM} corresponds to the tree-level vacuum energy difference for the SM. We also note that 
the contributions to $\Delta V_1$ from the gauge bosons $W$ and $Z$ and the top quark are identical in the 
SM and 2HDM, and so drop out from $\Delta \mathcal{F}_0$.
Combining \eqref{F_0} and \eqref{F_0_SM}, we obtain
\begin{eqnarray}	
\label{F_0_Final}
	\Delta\mathcal{F}_0 &=& - \frac{v^2}{8}c^2_{\beta-\alpha}\,(m_{H_0}^2 - m_h^2) 
	- \frac{m_h^4}{64\pi^2}\left( 3 + \log 2\right) 
	-\sum_{k} \frac{m_{0_k}^4}{64\pi^2}\left(\log\frac{|m_{0_k}^2|}{Q^2}-\frac{1}{2}\right) \\
	& + &  \frac{1}{64\pi^2} \sum_k \frac{1}{4}\left\{(v I_k)^2 - 2\,m_k^4 + \left[
	\left( v I_k - 2\,m_k^2 \right)^2 + m_k^2\,\left(v^2 J_k - v I_k \right) \right]\log\frac{m_k^2}{Q^2} \right\}\, , \nonumber
\end{eqnarray}
with $m_{0_k}^2$ the (possibly negative) squared scalar masses for $k = H^{\pm}, A_0, H_0, h$ evaluated at the origin. 
Further details on the derivation of $\Delta\mathcal{F}_0$ including explicit expressions for 
$I_k$ and $J_k$ are given in Appendix~\ref{app:HB_renorm}.
   
It is possible to show that the $Q^2$ dependence in~\eqref{F_0_Final} cancels out, so that $\Delta\mathcal{F}_0$ is 
renormalization scale independent. We also note that the first term in~\eqref{F_0_Final}, which corresponds to the tree-level 
contribution to $\Delta\mathcal{F}_0$, is negative definite and vanishes in the alignment limit $c_{\beta-\alpha} \to 0$. 
In this limit,~\eqref{F_0_Final} simplifies considerably and reads
\begin{eqnarray}
\label{F0_alignment}
	\Delta\mathcal{F}_0 &=& \frac{1}{64\,\pi^2}\left[ \left(m_h^2 - 2 M^2 \right)^2
		 \left(\frac{3}{2} +  
		\frac{1}{2}\, \mathrm{log}\left[ 
			\frac{4\,m_{A_0}\, m_{H_0} \, m^2_{H^{\pm}}}{\left(m_h^2 - 2 M^2\right)^2} \right]\right)        \right.\nonumber \\
 	&\ &+  \left. \frac{1}{2} \, \left( m^4_{A_0} +  m^4_{H_0} + 2\,m^4_{H^{\pm}} \right) + 
\left(m_h^2 - 2 M^2 \right) \left( m^2_{A_0} +  m^2_{H_0} + 2\,m^2_{H^{\pm}} \right)  \right].\,\,\,\,\,\,\,\,\,\,\,\,\,\,\,\,
\end{eqnarray}
%


\section{Vacuum Energy \textsl{vs} EW Phase Transition Strength:\\ Numerical scan}
\label{sec:scan2}

In order to show explicitly the correlation between the vacuum energy difference $\Delta\mathcal{F}_0$ and the nature of the EW 
phase transition in 2HDMs, we perform a Monte-Carlo scan over an extensive region of the 2HDM parameter space. We vary mass parameters from $100-1000$~GeV 
(but with $m_{H_0}>m_h$), and limit ourselves to the low $\tan\beta < 10$ region, since very large $\tan\beta$ is uninteresting for practical applications such 
as the baryon asymmetry computation. Each scanned point is tested for:
\begin{itemize}
	\item Tree-level unitarity and perturbativity (by requiring the tree-level quartic self-couplings among the \emph{physical} scalars to 
	be smaller than 2$\pi$)\footnote{In the literature, perturbativity is typically imposed as $\lambda_{1-5}<4\pi$. However, the scalar 
	vertex entering a loop expansion involves the self-coupling of \emph{physical} states, rather than the flavour eigenstates, hence the limits 
	must be imposed on the physical quartic couplings. Furthermore, we chose a more stringent upper bound of $2\pi$ for the tree-level couplings, as 
	this tends to ensure well-behaved running up to or beyond $\Lambda \gtrsim 2$~TeV. For the impact of requiring the \emph{running couplings} to 
	remain small all the way up to a certain cutoff scale, see discussion in section~\ref{sec:scan}.}.
	
	\item Stability of the electroweak vacuum at tree-level (c.f. eqs.~(\ref{eq:vacstability}) and (\ref{panicvacuum})) and at 1-loop level by directly 
	searching for lower secondary minima and/or unboundedness of the effective potential
up to a cutoff $\Lambda=5$~TeV\footnote{This is generally more stringent than evaluating the stability conditions in eq.~(\ref{eq:vacstability}) with the 
1-loop running couplings, as the latter method only takes the logarithmic contributions into account. Note also that one would find even more accurate 
exclusion regions by scanning the RG improved 1-loop effective potential with the 2-loop running couplings.}.
	\item Limits from EW precision observables~\cite{Grimus:2007if,Grimus:2008nb, Haber:2010bw, Funk:2011ad}.
	\item Flavour constraints, of which the most relevant in the low $\tan\beta$ region are $B^0-\bar{B^0}$ mixing~\cite{Geng:1988bq,Mahmoudi:2009zx} 
	and $\bar{B}\to X_s \gamma$ 
decays~\cite{Ciuchini:1997xe, Borzumati:1998tg, Ciafaloni:1997un, Hermann:2012fc, Misiak:2015xwa}.
	\item Bounds from direct scalar searches using \textsc{HiggsBounds}~\cite{Bechtle:2013wla}, and agreement with measured properties 
	of the $m_{h} = 125$~GeV Higgs boson  
using \textsc{HiggsSignals}~\cite{Bechtle:2013xfa}. 
\end{itemize}
A point passing all these tests is considered \emph{physical}. For each of these, the strength of the phase transition is computed by increasing 
the temperature, starting at $T=0$, and following the electroweak minimum (whose norm at temperature $T$ is denoted $v_T$), until we reach the critical 
temperature $T_c$ for which $\mathcal{F}_{T_c} =0$. The phase transition is considered strong if 
\begin{equation}
	\xi \equiv \frac{v_{T_c}}{T_c} \geq 1.
\end{equation}

\begin{figure}[h!]
	\centering
	\begin{picture}(440,175)
		\includegraphics[width=0.48\textwidth]{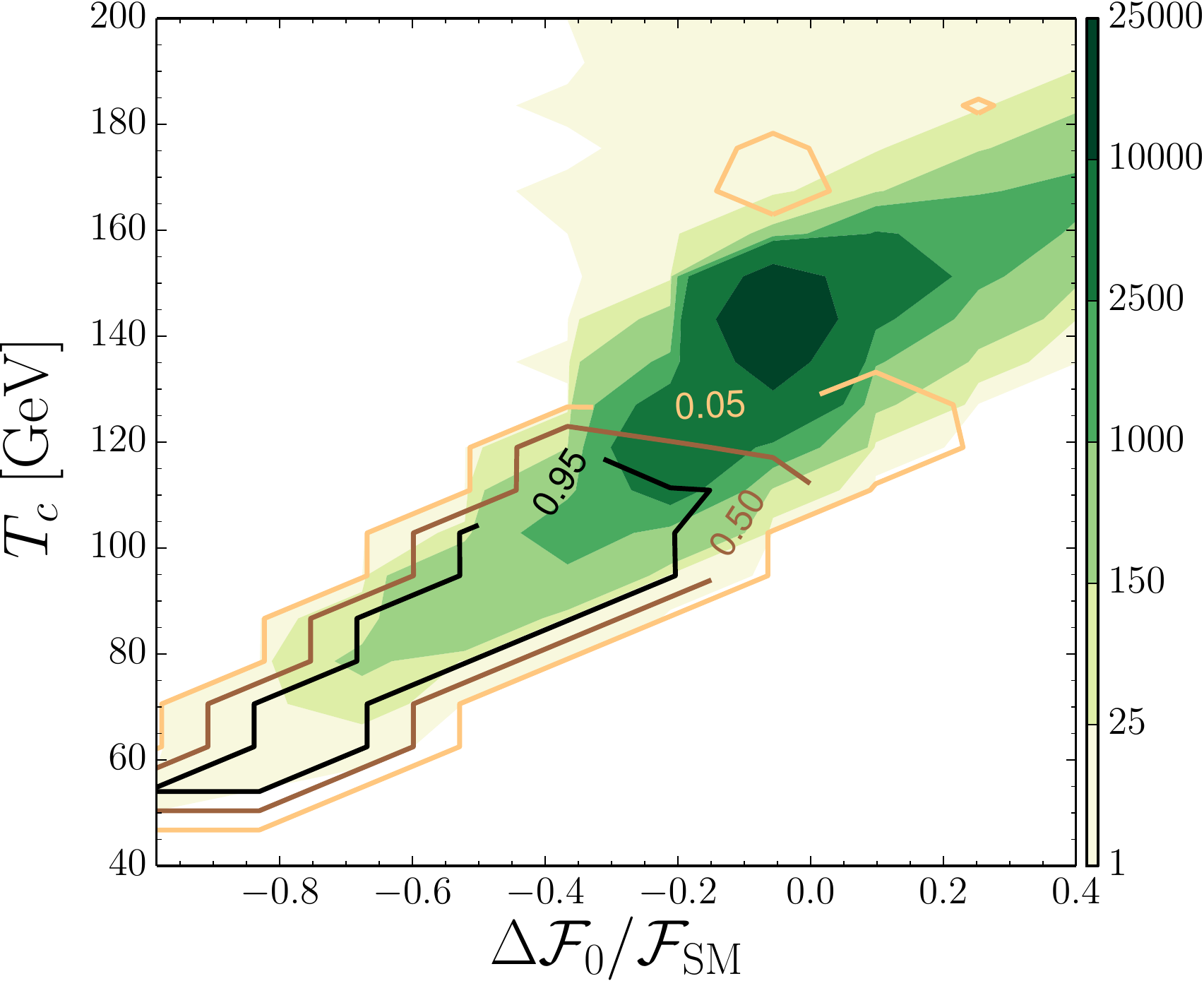}
		\qquad
		\includegraphics[width=0.48\textwidth]{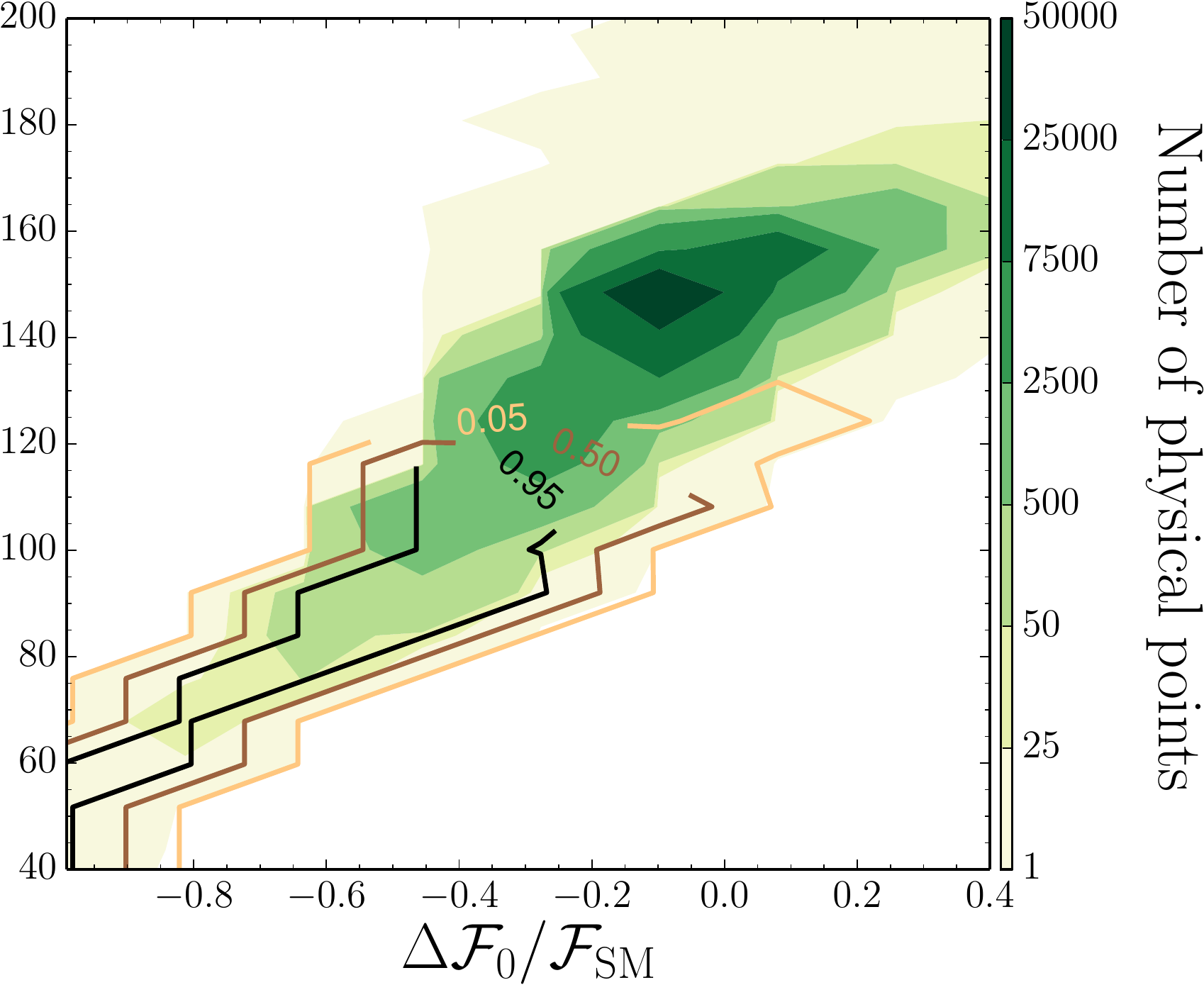}
		\put(-410,140){Type I}
		\put(-185,140){Type II}
	\end{picture}
	\\[-5mm]
	\begin{picture}(440,175)
		\includegraphics[width=0.48\textwidth]{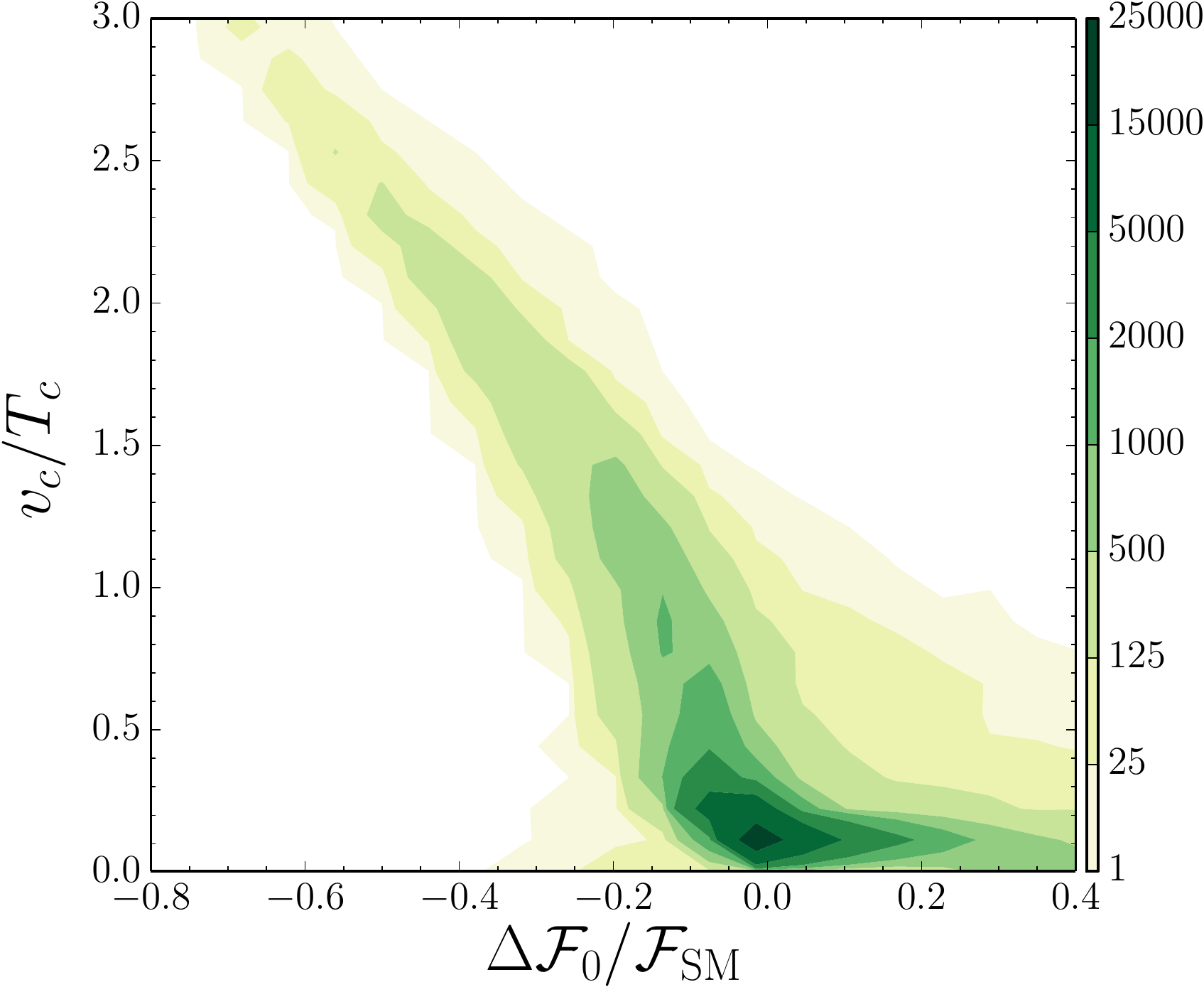}
		\qquad
		\includegraphics[width=0.48\textwidth]{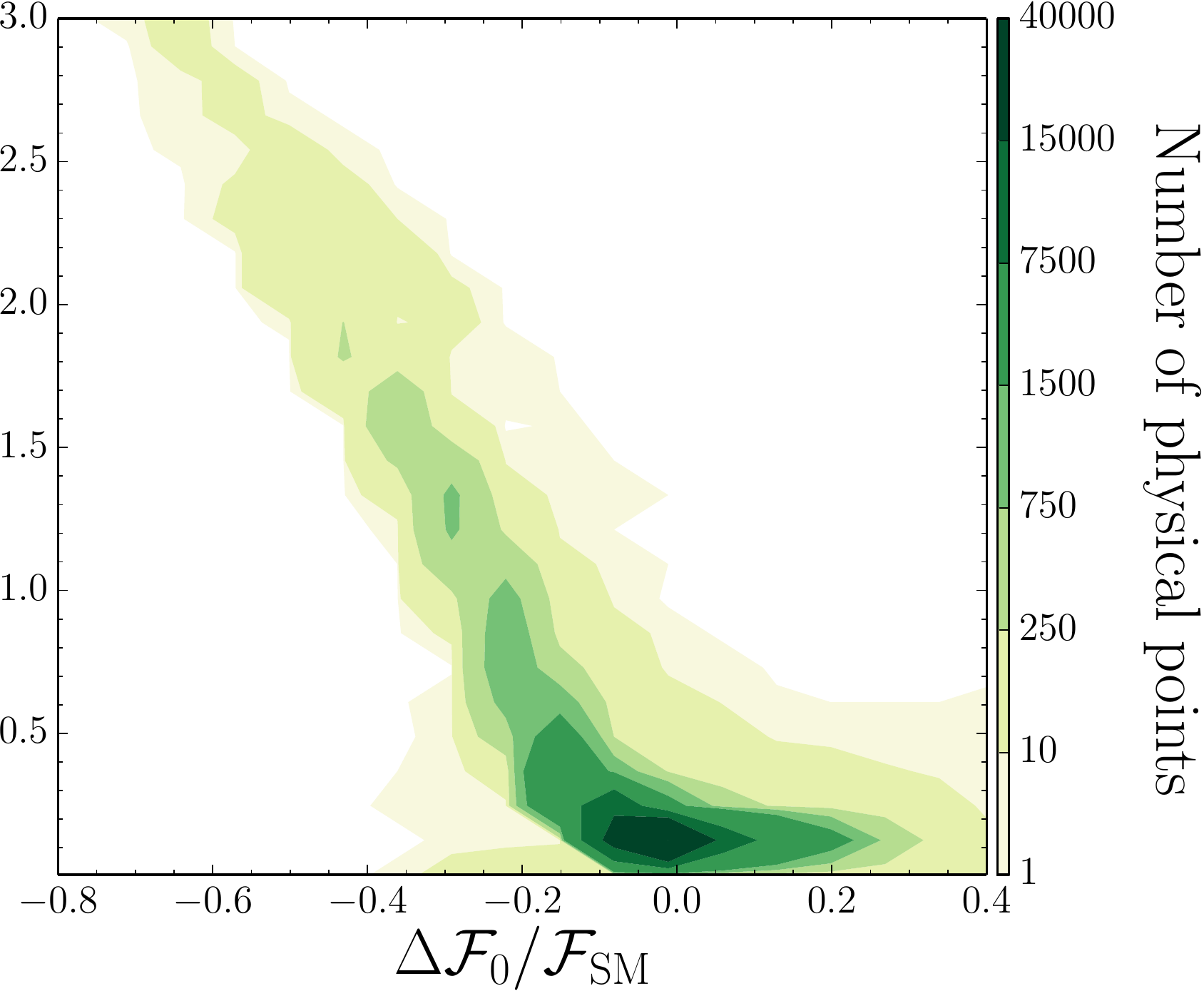}
		\put(-325,140){Type I}
		\put(-100,140){Type II}
	\end{picture}
	\caption{\small Results of a numerical scan of the 2HDM parameter space (see text for details) showing the correlation between the $\Delta\mathcal{F}_0$ and (top) the critical temperature (bottom) the strength of the 
	EWPT for Type I (left) and Type II (right). Filled contours indicate the density of physical points. Also shown are contours of $\mathcal{P}_{\xi>1}$, the posterior probability of having a strong first order EWPT.}
	\label{Figure_Correlation}
\end{figure}

Clearly, the larger $\Delta\mathcal{F}_0$ is, the smaller the temperature corrections required in order to reach $\mathcal{F}_{T_c} =0$. 
Since $v_T$ also grows as $T$ decreases, the overall result is that the strength of the phase transition should be directly related to $\Delta\mathcal{F}_0$. 
This is illustrated in Fig.~\ref{Figure_Correlation}. Here, the filled green contours indicate the number of physical points in a given region of the parameter 
space. In any such region we also define
\begin{equation}
	\label{P_xi}
	\mathcal{P}_{\xi>1} \equiv \frac{\text{\# points with $\xi>1$}}{\text{\# physical points}},
\end{equation}
whose contours are shown in the empty curves indicating the percentage of points in the encircled region for which the phase transition 
is strong (e.g. in Fig.~\ref{Figure_Correlation} (top), $95\%$ of points inside the black solid curve have $\xi \geq 1$). Note that the latter curves, 
being the ratio of density distributions in a certain region, are less sensitive to the priors of the scan than the actual distribution of points alone, 
and therefore offer a more meaningful physical picture in that they can be interpreted as a posterior probability density for requiring a 
strongly first-order EWPT given the existing constraints on the model.

For convenience, we normalize the vacuum energy by the SM value at 
1-loop\footnote{As $\mathcal{F}_0^{\rm SM}$ is negative, larger values of $\Delta \mathcal{F}_0$ will correspond to more negative 
values of $\Delta \mathcal{F}_0/\mathcal{F}_0^{\rm SM}$.} \mbox{$\mathcal{F}_0^{\rm SM}\approx-1.25\times 10^8~\text{GeV}^4$}. 
It is clear from Fig.~\ref{Figure_Correlation} (top) that as $\Delta\mathcal{F}_0/\mathcal{F}_0^{\rm SM}$ decreases both
$T_c$ and the likelihood of having a strong phase transition increase. Notice, furthermore, 
that the phase transition is guaranteed to be strong if $\Delta \mathcal{F}_0/\mathcal{F}_0^{\rm SM}\lesssim -0.34$ for the sample generated in our scan. This 
can be used as an efficient criterion to judge the nature of the phase transition, as it does not require the evaluation of the thermal potential (although it is not used 
in what follows). 
We however emphasize that the details of the temperature-dependent part of the effective potential are obviously important for the thermal 
evolution of the system, and oftentimes one cannot precisely judge the nature of the phase transition by the vacuum energy alone. E.g. for $\Delta \mathcal{F}_0=0$ in Type I, the EWPT can be weak or strong, as shown in Fig.~\ref{Figure_Correlation} (bottom, left).

\begin{figure}[h!]
	\centering
	\begin{picture}(440,175)
		\includegraphics[width=0.48\textwidth]{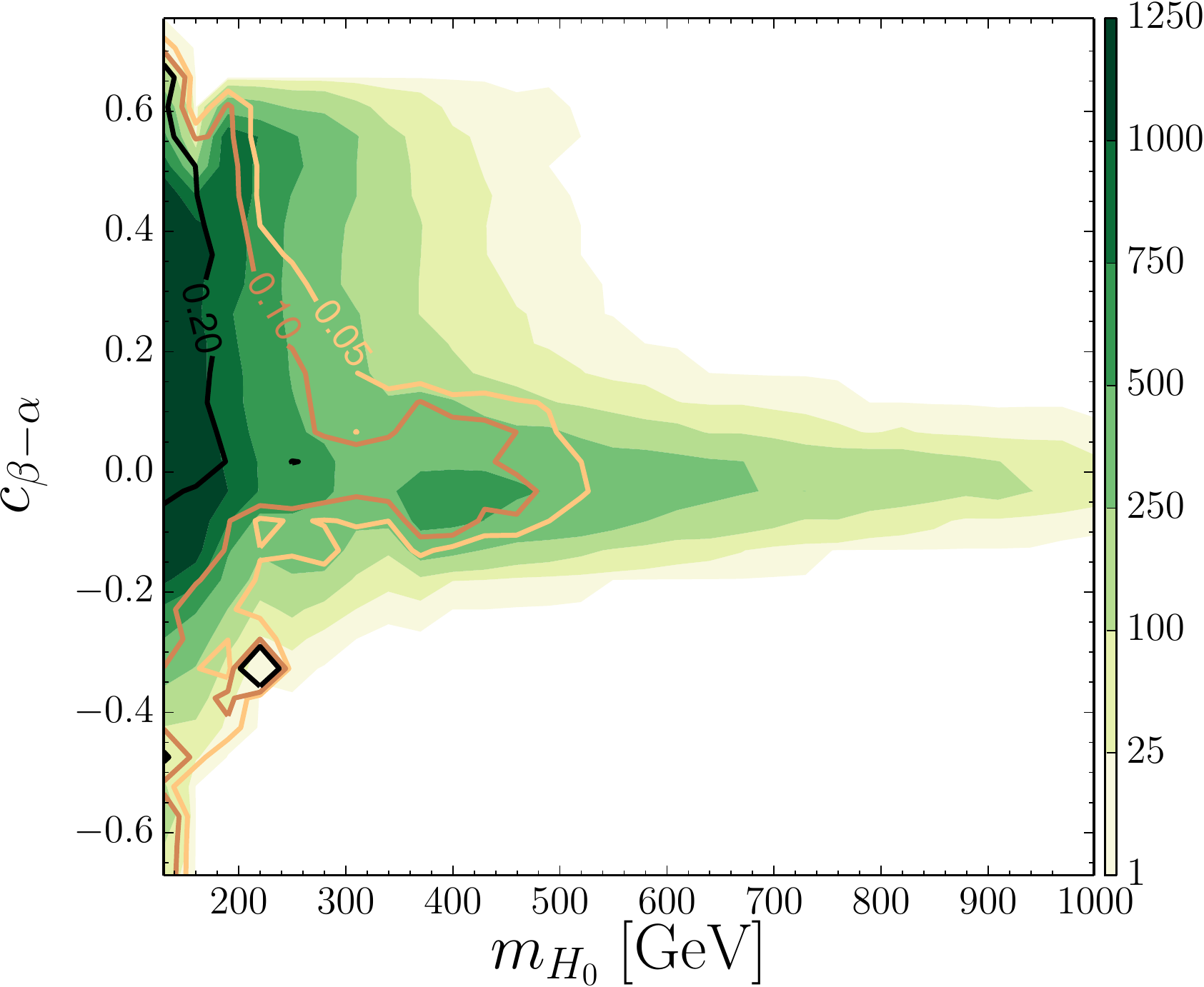}
		\qquad
		\includegraphics[width=0.48\textwidth]{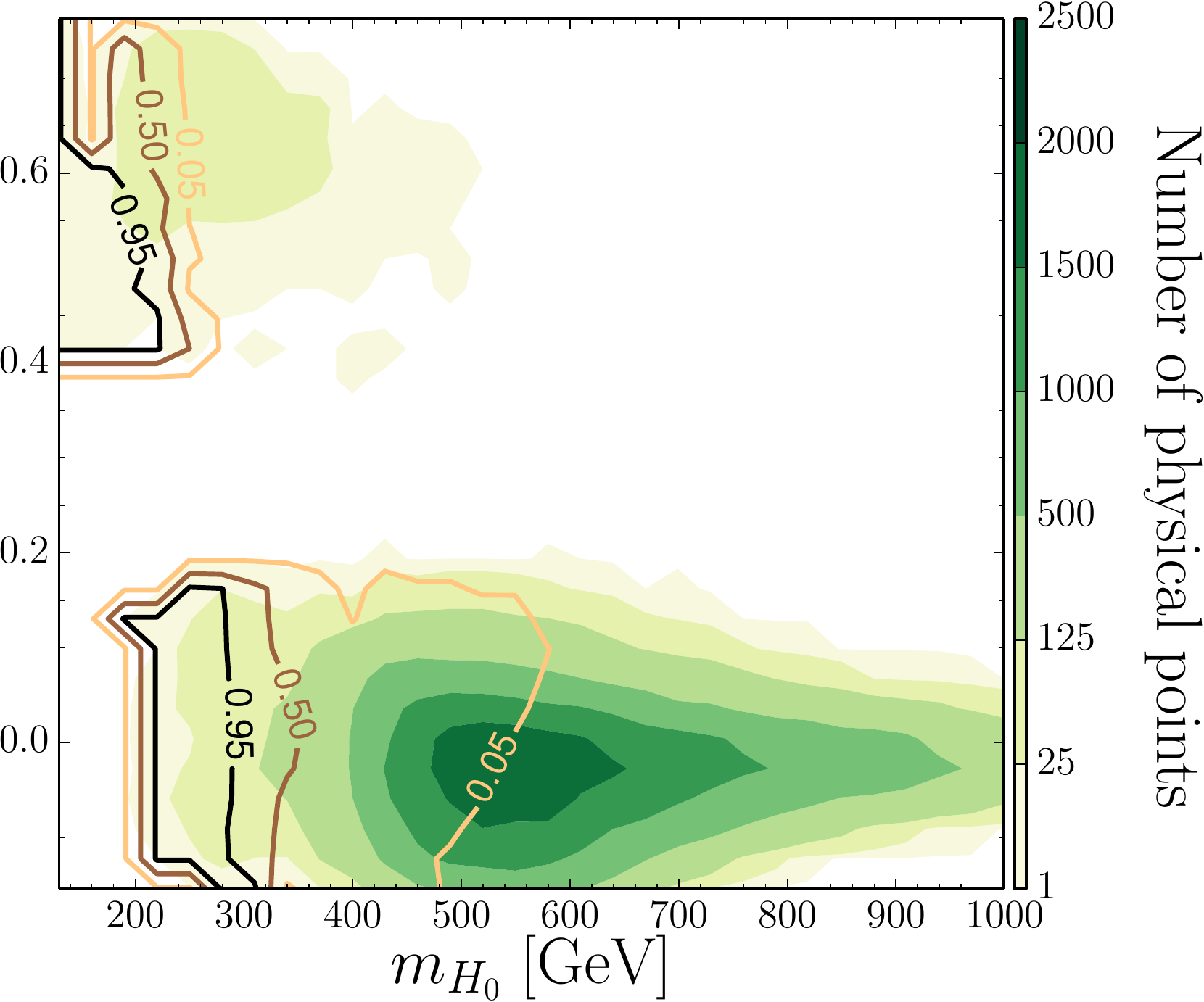}
		\put(-310,135){Type I}
		\put(-90,135){Type II}
	\end{picture}
	\caption{Distribution of physical points, as in Figure~\ref{Figure_Correlation}, and $\mathcal{P}_{\xi>1}$ contours in the $(m_{H_0},\, c_{\beta-\alpha})$ plane. As $H_0$ gets heavier, a strong first order EWPT increasingly favours alignment. In Type II the wrong-sign scenario, albeit less populated, can also lead to a strong EWPT.}
	\label{fig:abxmH}
\end{figure}

Yet, a direct correlation certainly exists between these quantities, from which one can understand and predict the favoured corners of the parameter 
space for a strong EWPT. Eq.~(\ref{F_0_Final}) shows that the vacuum energy difference receives a negative tree-level 
contribution away from alignment, which increases with $m_{H_0}$. We thus expect a strong EWPT to favour 
the alignment limit, and the more so the heavier $H_0$ is. 
These expectations are confirmed by the data, as shown in Fig.~\ref{fig:abxmH}. In both Type I and II scenarios the probability contours 
increasingly favour alignment for a strong EWPT as $m_{H_0}$ grows. For Type I, even though the distribution of 
physical points already narrows around alignment for $m_{H_0}\gtrsim 550$~GeV, the narrowing of the $\mathcal{P}_{\xi>1}$ bands is significantly  
more drastic and does not merely follow that of the physical distribution. It is also worth noticing that, while for Type I the low-mass region 
is the mostly populated, for Type II the lower bound $m_{H^\pm}>480$~GeV from flavour constraints tends to shift the masses of the additional 
scalars towards rather large values, which is why the physical points are mostly concentrated in the region of $m_{H_0}\sim 500$~GeV. 
For Type II we also note the physical region for $c_{\beta-\alpha} \gtrsim 0.4$, corresponding to the 2HDM wrong-sign scenario~\cite{Ferreira:2014naa}. 
Both in Type I and II scenarios one sees that away from the alignment limit 
there is a tension between a strong EWPT and a heavy $H_0$.

\begin{figure}[t!]
	\centering
	\begin{picture}(420,175)
		\includegraphics[width=0.48\textwidth]{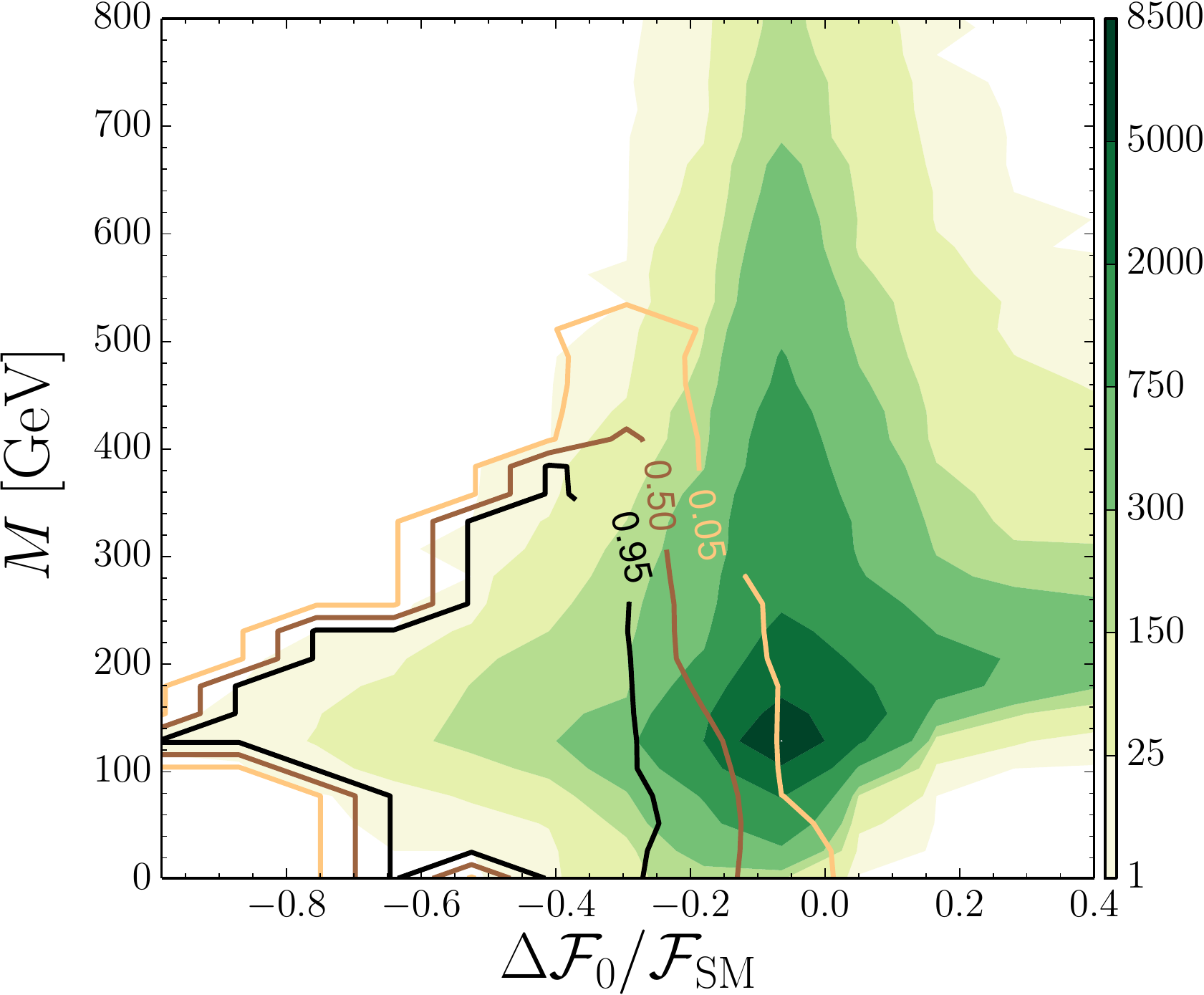}
		\qquad
		\includegraphics[width=0.48\textwidth]{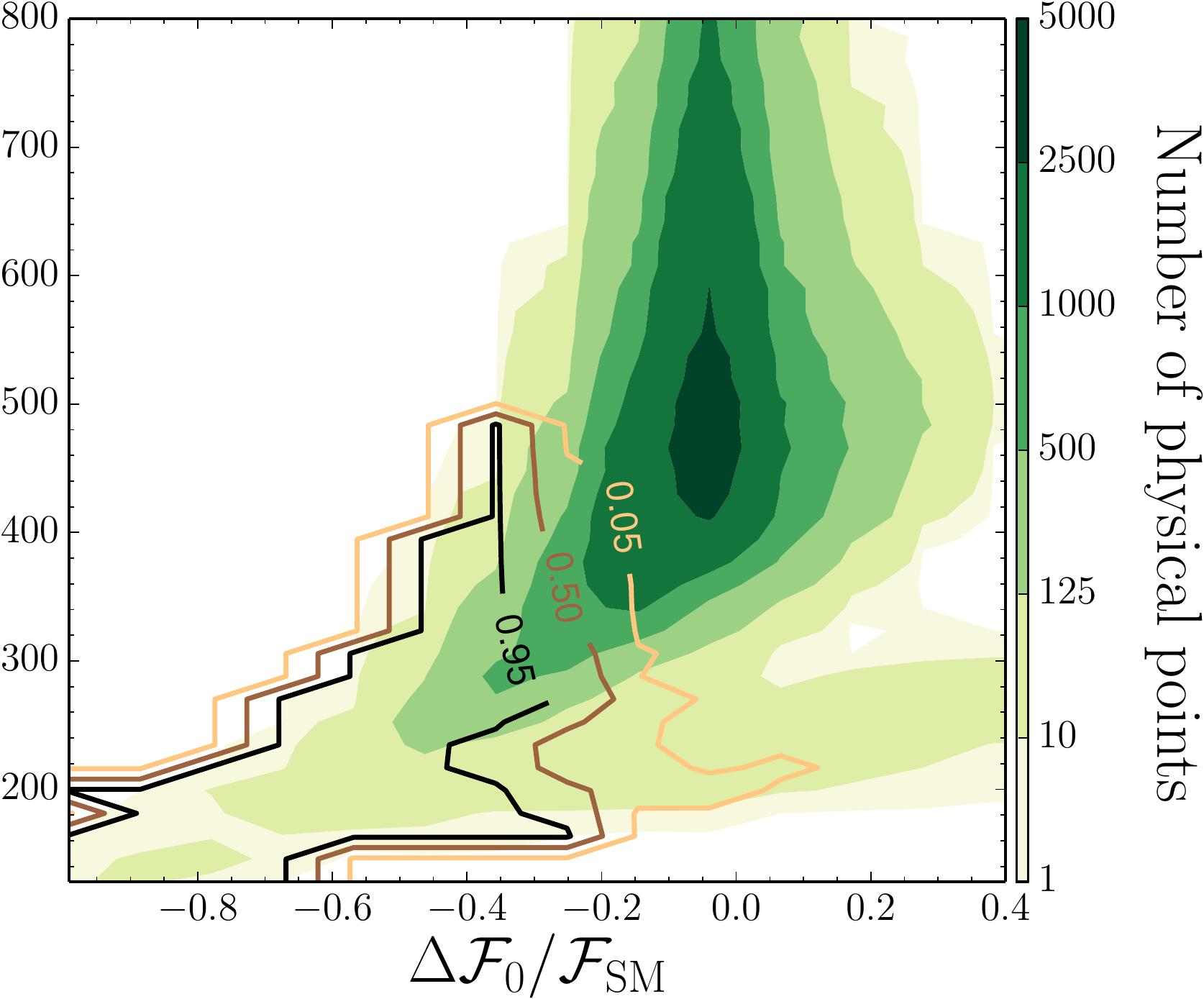}
		\put(-395,140){Type I}
		\put(-175,140){Type II}
	\end{picture}
	\caption{Distribution of physical points, as in Figure~\ref{Figure_Correlation}, and $\mathcal{P}_{\xi>1}$ contours in the $(\Delta\mathcal{F}_0/\mathcal{F}_0^{\mathrm{SM}},\,M)$ plane.}
	\label{fig:MxdV}
\end{figure}

The dependence of the vacuum uplifting with the overall mass scale $M$ is determined mostly by stability, perturbativity and unitarity constraints. 
Indeed, close to the alignment limit the quartic couplings $\lambda_{1,2}$ read
\be
\label{l1l2}
v^2\,\lambda_1 \approx m_h^2 +  t^2_\beta\, \Omega^2\,\,\,,\quad  v^2\,\lambda_2 \approx m_h^2 +  t^{-2}_\beta\, \Omega^2,
\ee
where the parameter
\be
	\label{Omega}
	\Omega^2 \equiv m_{H_0}^2 - M^2
\ee 
has been introduced for its usefulness in the analysis of the stability and unitarity requirements. 
Recalling eq.~\eqref{eq:vacstability}, both couplings $\lambda_{1,2}$ must be positive and it follows that
\be
	m_h^2 > -{\rm max}(t^2_\beta,\ t^{-2}_\beta)\,\Omega^2,
\ee
so that as $M^2$ grows larger, $m_{H_0}^2$ has to follow it closely. In addition eq.~\eqref{eq:vacstability} shows that
\be\begin{split}
	v^2\,\lambda_3 & \approx 2 m_{H^\pm}^2 - 2 m_{H_0}^2 + \Omega^2 + m_h^2,\\
	v^2\,\lambda_4 & \approx m_{A_0}^2 - m_{H_0}^2 + \Omega^2 - 2 m_{H^\pm}^2
\end{split}\ee
cannot grow too negative either, from which it follows that $m_{H^\pm}^2$ and $m_{A_0}^2$ cannot be much smaller than a large $M^2$. In summary, for $M^2 \gg m_h^2$, 
stability enforces $m^2_{H_0}, m^2_{A_0}, m^2_{H^\pm} \sim M^2$, for which the decoupling limit is approached and $\Delta\mathcal{F}_0 \to 0$, as can be verified 
by setting $m_{H_0} = m_{H^\pm} = m_{A_0} \approx M \gg m_h$ in eq.~(\ref{F_0_Final}). Therefore, a significant uplifting of the vacuum energy can only be 
achieved for $M\sim v$, which is confirmed by Fig.~\ref{fig:MxdV}. We note that again in Type II the distribution of physical points is peaked 
around larger values due to the lower bound on $m_{H^\pm}$ from the $\bar{B}\to X_s \gamma$ constraint. 
However, in both types a moderate uplifting of the vacuum energy is achieved only for $M\lesssim 500$~GeV.

\begin{figure}[h!]
	\centering
	\begin{picture}(420,175)
		\includegraphics[width=0.48\textwidth]{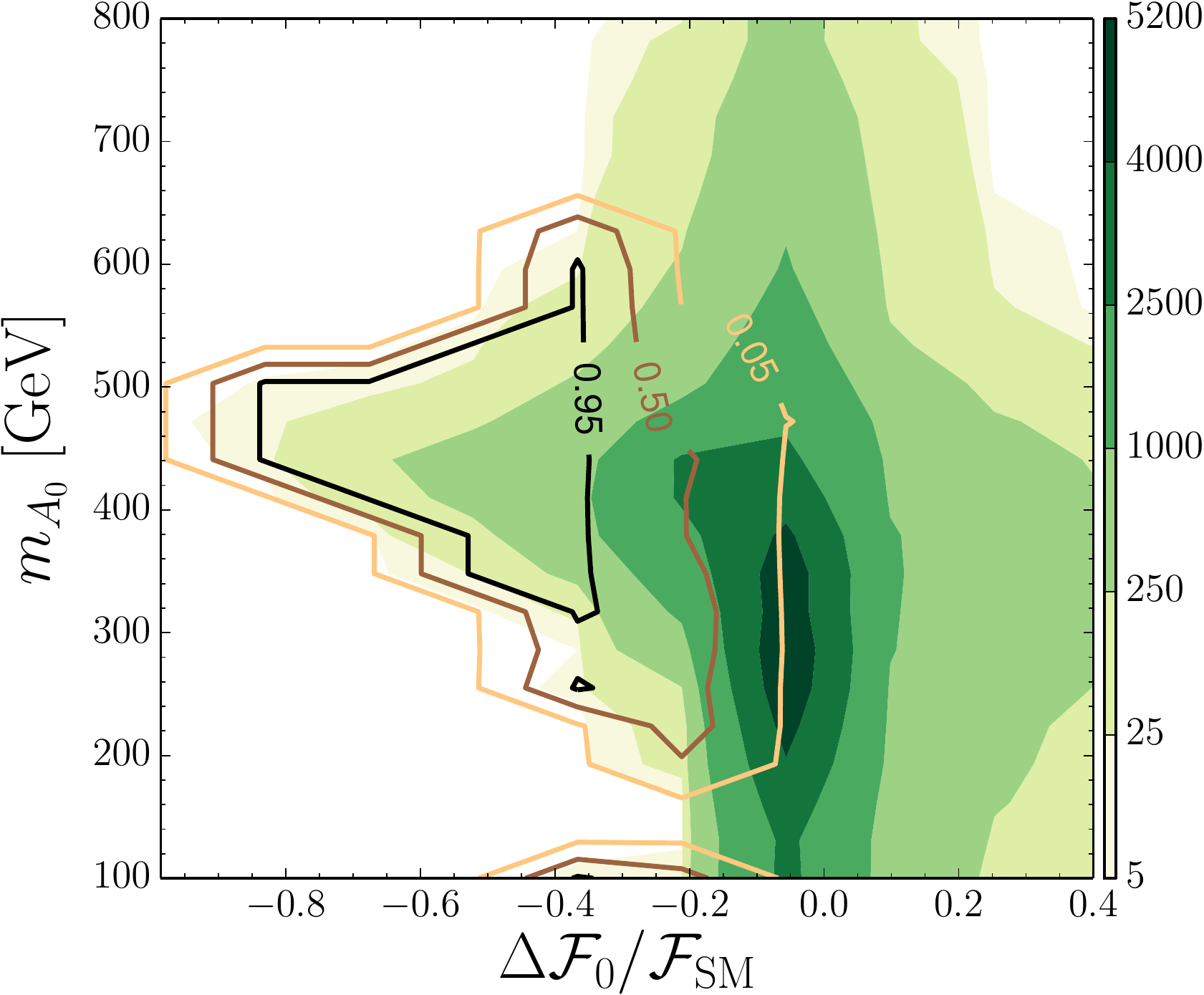}
		\qquad
		\includegraphics[width=0.48\textwidth]{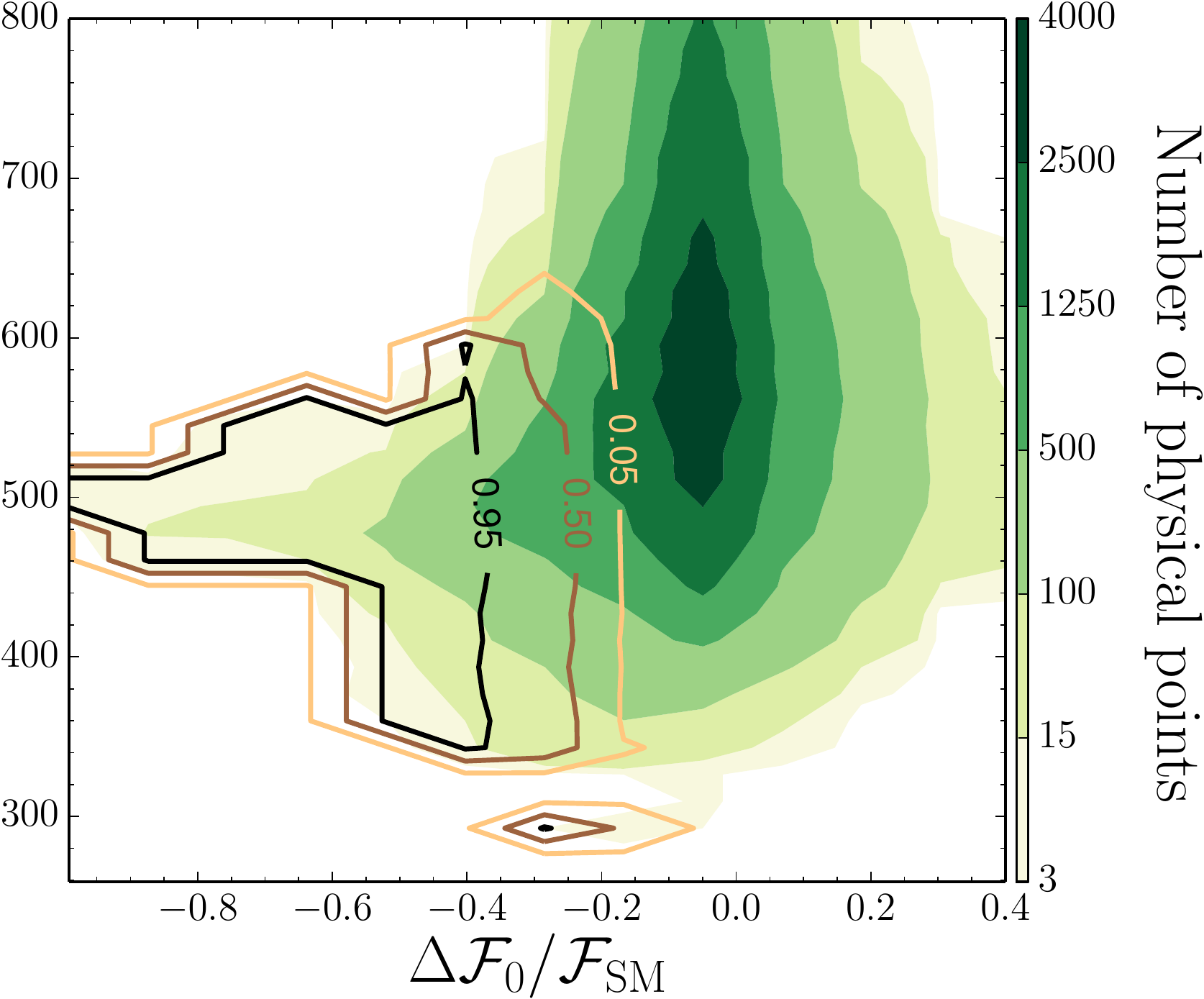}
		\put(-390,140){Type I}
		\put(-170,140){Type II}
	\end{picture}
	\caption{Distribution of physical points and $\mathcal{P}_{\xi>1}$ contours in the $(\Delta\mathcal{F}_0/\mathcal{F}_0^{\mathrm{SM}}, \,m_{A_0})$ plane.}
	\label{fig:mChxdV}
\end{figure}

A strongly first order EWPT generally relies on the existence of sizable couplings between the symmetry breaking scalar field (the Higgs) and the particles 
in the plasma, which means that one or more of the additional scalars must be significantly heavier than the overall 
mass scale $M$, as the mass splitting would be controlled by these large couplings. 
We have already established that a large $m_{H_0}$ becomes disadvantageous for a strong EWPT away (even if only 
slightly) from alignment. Furthermore, for $t_{\beta} \neq 1$ a large $\Omega^2$ quickly violates perturbativity bounds. On the other hand, 
EW precision observables constrain the charged scalar $H^{\pm}$ to be close in mass to either $m_{H_0}$ or $m_{A_0}$. This leaves $A_0$ as the 
only scalar whose mass is free to be large\footnote{$H^{\pm}$ may also be significantly heavier than $M$ if paired to $A_0$, but not on its own.}, 
and Fig.~\ref{fig:mChxdV} confirms that a rather heavy\footnote{We note that a heavy pseudoscalar ($m_{A_0}^2 \gg M^2$) does induce a negative quartic coupling 
$\lambda_5 = (M^2 - m_{A_0}^2)/v^2$. However, this does not pose a problem for stability, since only the absolute value of $\lambda_5$ 
enters eq.~(\ref{eq:vacstability}).} $A_0$ is indeed the most favoured scenario, with $> 94$\% of strong 
phase transition points lying above the lower bound $m_{A_0}\gtrsim 300$~GeV. 

\begin{figure}[h!]
	\centering
	\begin{picture}(420,175)
		\includegraphics[width=0.48\textwidth]{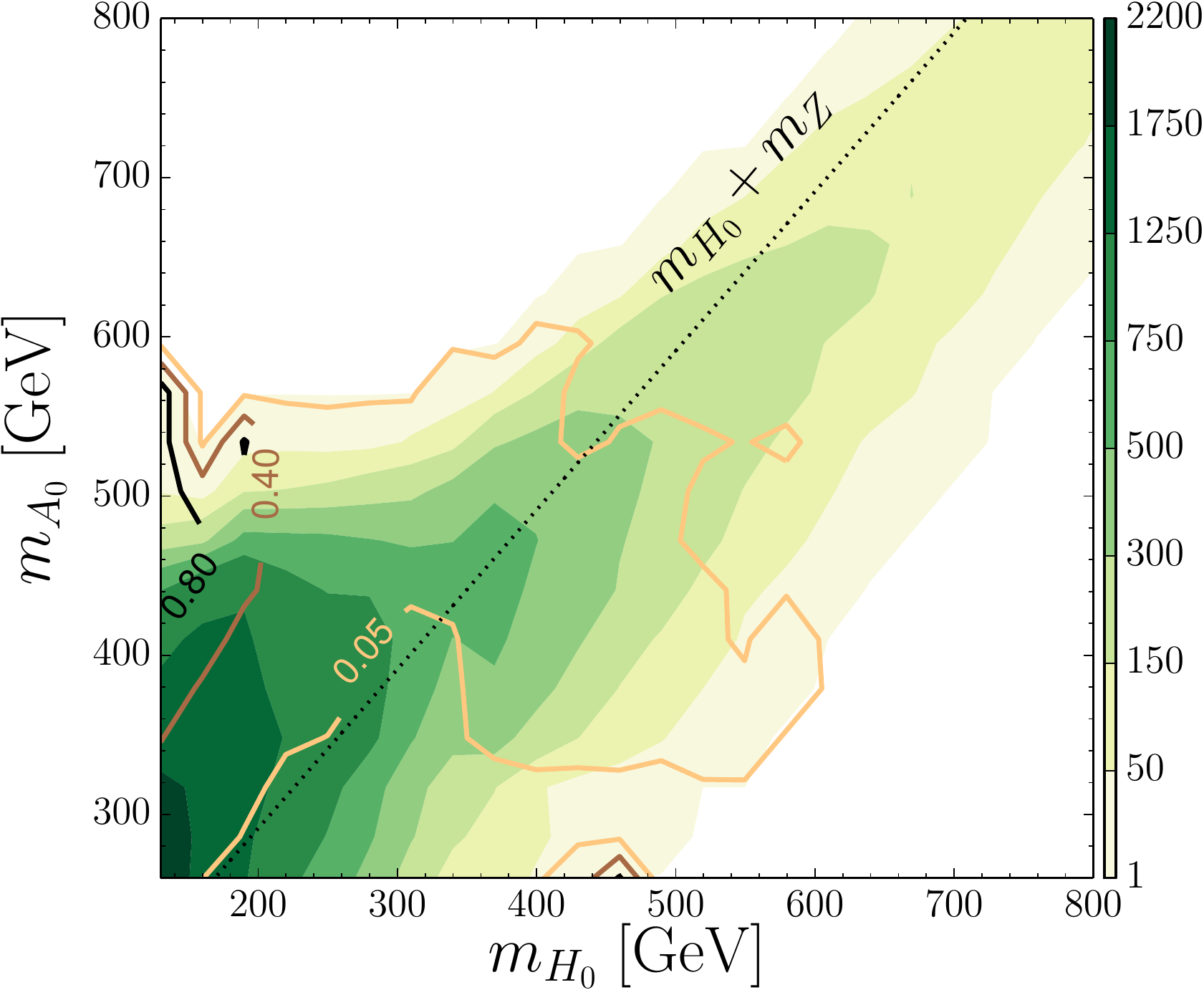}
		\qquad
		\includegraphics[width=0.48\textwidth]{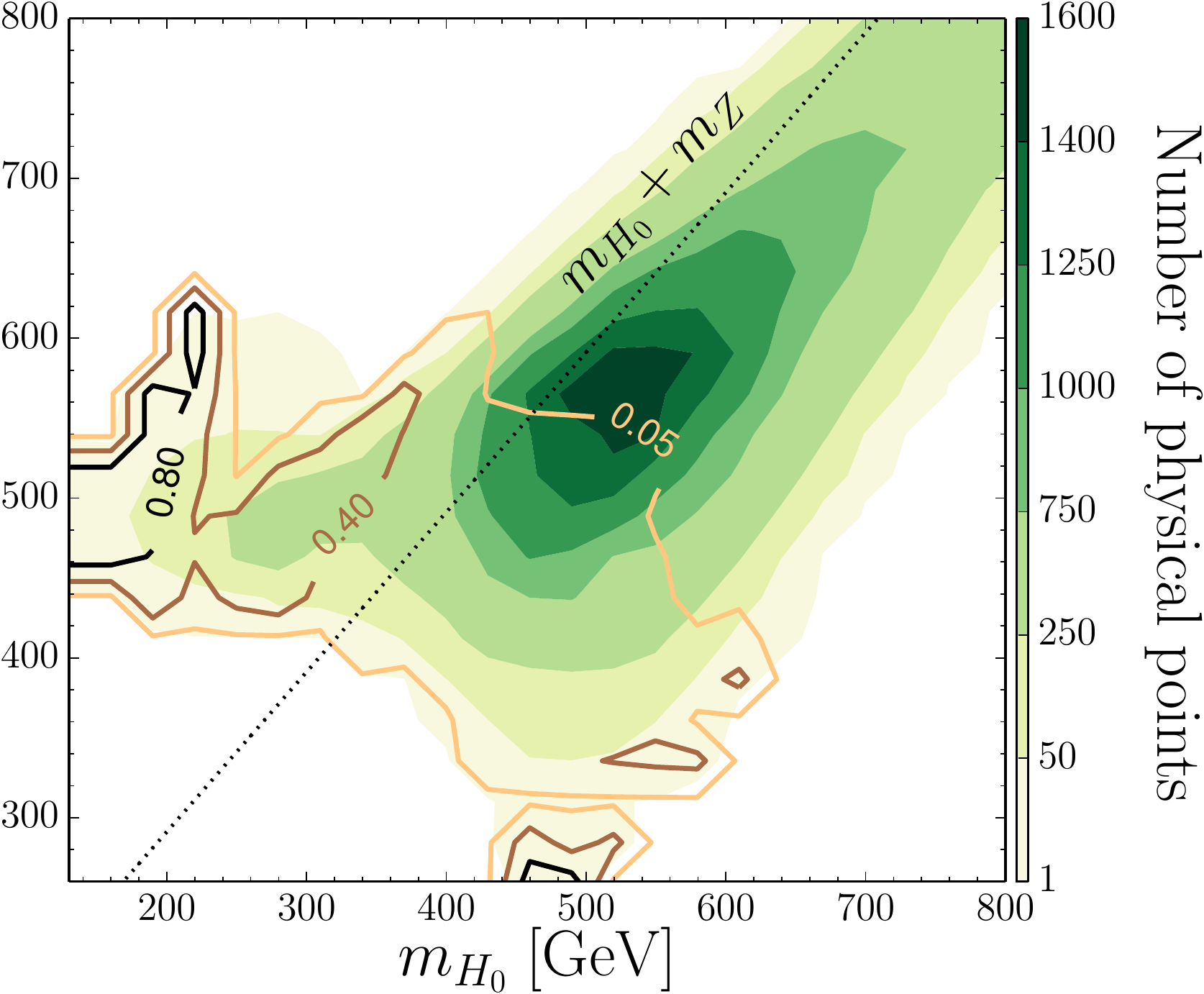}
		\put(-390,140){Type I}
		\put(-170,140){Type II}
	\end{picture}
	\caption{Distribution of physical points and $\mathcal{P}_{\xi>1}$ contours in the ($m_{H_0},\, m_{A_0}$) plane. A strong first order EWPT is clearly favoured by a splitting $m_{A_0}> m_{H_0}+m_Z$.}
	\label{fig:mAxmH}
\end{figure}

These results are put together in Fig.~\ref{fig:mAxmH}, illustrating how the likelihood of a strong EWPT varies with $m_{H_0}$ and $m_{A_0}$. 
In both Type I and II 2HDM scenarios a strong transition favoures a large splitting $m_{A_0} > m_{H_0}+m_Z$, pointing to the $A_0\to Z H_0$ 
decay as a \emph{smoking gun signature} of a 2HDM with a strongly first order EWPT. 
The detection prospects of this channel, and its importance as complementary to searches into SM final states, have been discussed 
in~\cite{Dorsch:2014qja,Coleppa:2014hxa,Kling:2016opi,Dorsch:2016tab}.

\section{Analytic results}
\label{sec:scan}

We now turn to an analytic exploration of the 2HDM vacuum uplifting 
as computed from eq.~(\ref{F_0_Final}).~Given the large dimensionality of the 2HDM parameter space, we perform the study 
in various limits which allow us to explicitly investigate the relevant parameter dependences.  
In the following section we focus on the alignment limit, pair $m_{H^\pm}$ exactly with 
either $m_{H_0}$ or $m_{A_0}$, and work out the dependence of the vacuum energy and phase transition strength with the 
splitting $\Delta m_{\mathrm{AH}} \equiv m_{A_0} - m_{H_0}$ and $\Omega \equiv \sqrt{\left|\Omega^2\right|}\times\mathrm{sign}(\Omega^2)$ for 
different fixed values of $m_{H_0}$.  Then, in section~\ref{2HDM_NA} we allow for deviations from the alignment limit, fixing a degenerate 
spectrum ($m_{H_0}=m_{A_0}=m_{H^\pm}$) for simplicity. Finally we devote section~\ref{sec:I2HDM} to the special case of the Inert 2HDM where only one double takes a vev and the $\mathbb{Z}_2$ symmetry is exact.

\subsection{The Alignment Limit $c_{\beta-\alpha} = 0$}
\label{Sec_alignment}

We start by considering the 
alignment limit $c_{\beta-\alpha} = 0$, where $h$ behaves exactly as the SM Higgs boson.
In this case, $\Delta\mathcal{F}_0$ is given by~\eqref{F0_alignment}. Since measurements of EW precision observables 
(in particular the $T$-parameter) require an approximate degeneracy $m_{H^{\pm}} \sim m_{H_0}$ or $m_{H^{\pm}} \sim m_{A_0}$, we set 
for simplicity this pairing as exact, analysing both possibilities. With these parameters fixed, $\Delta\mathcal{F}_0$ is then solely 
dependent on $m_{H_0}$,  $m_{A_0}$, and $\Omega^2$.

\begin{figure}[t]
\centering

$\vcenter{\hbox{\includegraphics[width=0.385\textwidth]{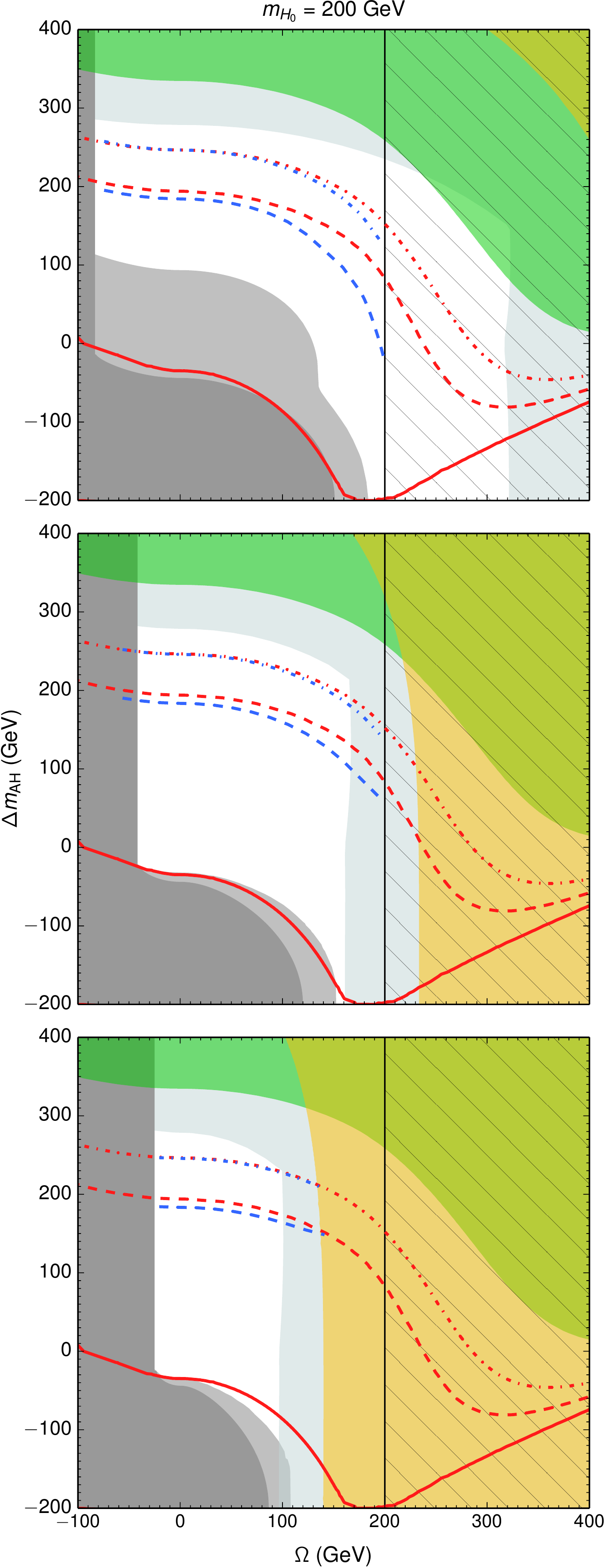}}}$
$\vcenter{\hbox{\includegraphics[width=0.385\textwidth]{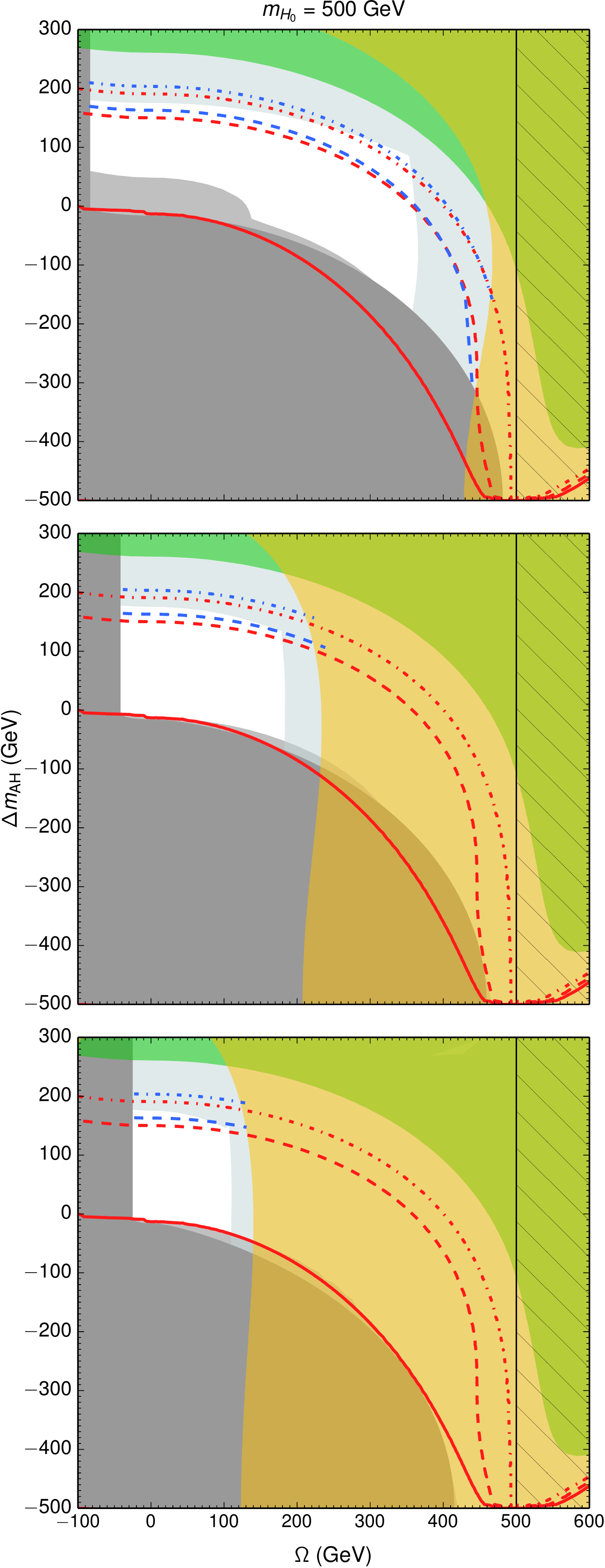}}}$
$\vcenter{\hbox{\includegraphics[width=0.21\textwidth]{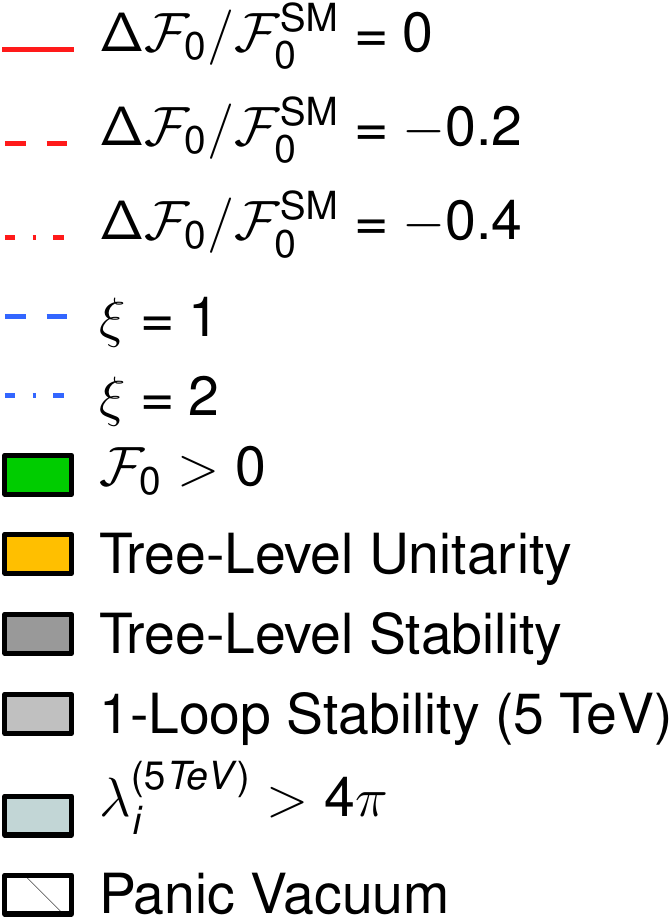}}}$

\caption{\small{$\Omega \equiv \sqrt{\left|\Omega^2\right|}\times\mathrm{sign}(\Omega^2)$ vs 
$\Delta m_{\mathrm{AH}} \equiv m_{A_0} - m_{H_0}$ assuming $m_{H^{\pm}} = m_{A_0}$, for 
$m_{H_0} = 200,\,500$ GeV (Left to Right) and $t_{\beta} = 1.5,\,3,\,5$ (Top to Bottom).
Red lines show constant values of $\Delta\mathcal{F}_0/\mathcal{F}^{\mathrm{SM}}_0$. 
Blue lines show constant values of the strength of the EWPT $\xi$. 
The grey region is excluded by boundedness from below of the scalar potential, while the brown region is excluded 
by unitarity. In the hatched region, a panic vacuum develops.\label{Vacuum_Energy_Alignment_1}}}
\end{figure}

We first fix $m_{H^{\pm}} = m_{A_0}$ and show in Fig.~\ref{Vacuum_Energy_Alignment_1} 
the parameter space regions of constant $\Delta\mathcal{F}_0/\mathcal{F}^{\mathrm{SM}}_0$
in the ($\Omega$, $\Delta m_{\mathrm{AH}}$) plane,
respectively for $m_{H_0} = 200,\,500$ GeV (Left to Right) and $t_{\beta} = 1.5,\,3,\,5$ (Top to Bottom).
In each case we show the constraints from tree-level unitarity, boundedness from below of the scalar potential 
and non-existence of a panic vacuum. We note that as opposed to unitarity and stability, 
$\Delta\mathcal{F}_0/\mathcal{F}^{\mathrm{SM}}_0$ and the existence of a panic vacuum do not depend on $t_{\beta}$ (this last one for 
$c_{\beta-\alpha} = 0$).
To estimate the breakdown of perturbativity, we show the region for which any quartic coupling grows 
larger than $4\pi$ at a cutoff $\mu=5$~TeV from 2-loop running~\cite{Chowdhury:2015yja}, starting from $\mu_0 = {\rm max}(m_{H_0}, m_{H^\pm}, m_{A_0})$ to 
ensure that the heavy degrees of freedom will only contribute above their threshold. While this is not a hard limit on the model compared to the others 
presented, it provides an idea of the UV scale of new physics that would be required in such a picture. Finally, we also show the lines of a constant 
strength of the EWPT $\xi$
in the ($\Omega$, $\Delta m_{\mathrm{AH}}$) plane, obtained numerically. These smoothly track the lines of constant 
$\Delta\mathcal{F}_0/\mathcal{F}^{\mathrm{SM}}_0$, confirming the observations in section~\ref{sec:scan2}
regarding the tight correlation between the strength of the EWPT and $\Delta\mathcal{F}_0$ in 2HDM scenarios.

\begin{figure}[t]
\centering

$\vcenter{\hbox{\includegraphics[width=0.385\textwidth]{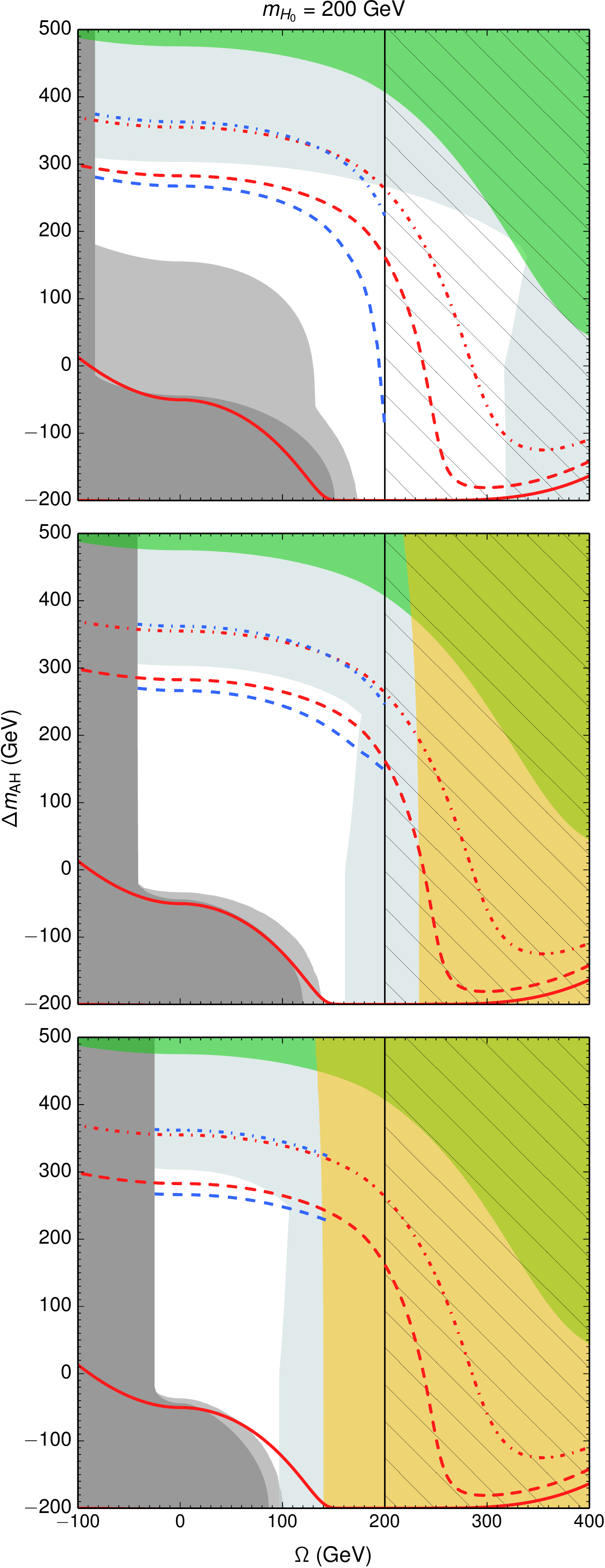}}}$
$\vcenter{\hbox{\includegraphics[width=0.385\textwidth]{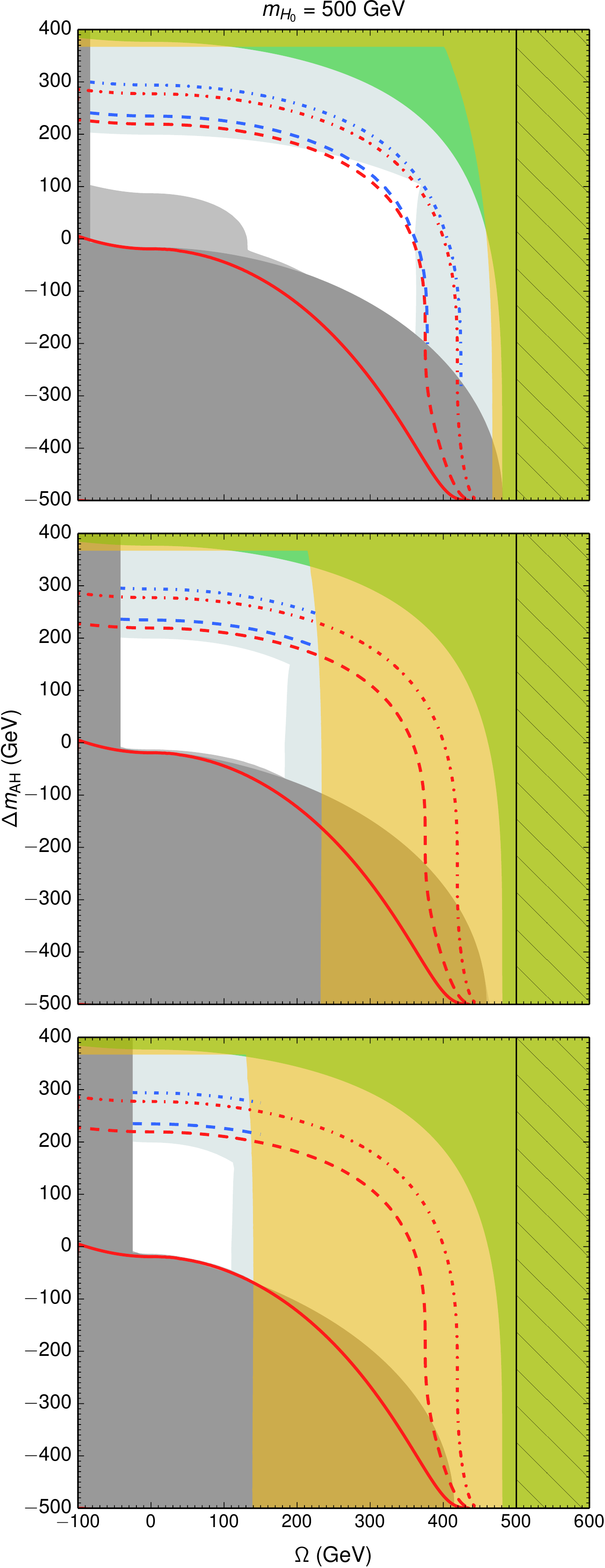}}}$
$\vcenter{\hbox{\includegraphics[width=0.21\textwidth]{Plots/Vacuum_Energy_Legend.pdf}}}$

\caption{\small{$\Omega$ vs 
$\Delta m_{\mathrm{AH}}$ assuming $m_{H^{\pm}} = m_{H_0}$, for 
$m_{H_0} = 200,\,500$ GeV (Left to Right) and $t_{\beta} = 1.5,\,3,\,5$ (Top to Bottom). Labels as in 
Fig.~\ref{Vacuum_Energy_Alignment_1}.\label{Vacuum_Energy_Alignment_2}}}
\end{figure}

From Fig.~\ref{Vacuum_Energy_Alignment_1} we see that a strongly first order EWPT is achieved by increasing $\Delta m_{\mathrm{AH}}$ 
in all cases. For $m_{H_0} \gg v$ ($m_{H_0} = 500$ GeV in Fig.~\ref{Vacuum_Energy_Alignment_1}) and $t_{\beta} \sim 1$ 
it is also possible to achieve such a strongly first order transition by increasing $\Omega$ (with $\Omega < m_{H_0}$) for 
$\Delta m_{\mathrm{AH}} < 0$, but this possibility is forbidden by unitarity as $t_{\beta}$ departs significantly from 1.
We repeat the analysis, now for $m_{H^{\pm}} = m_{H_0}$, and show the results in Fig.~\ref{Vacuum_Energy_Alignment_2}. These are qualitatively similar to those 
from Fig.~\ref{Vacuum_Energy_Alignment_1} for the $m_{H^{\pm}} = m_{A_0}$ scenario. Together, these show that  
a strongly first order EWPT within the 2HDM generically favours $m_{A_0} - m_{H_0} \gtrsim 100$ GeV, 
leading to the landmark signature $A_0 \to H_0 Z$ at colliders.

Before continuing, let us note that in our analytical study of the 2HDM vacuum energy 
we haven't imposed several experimental constraints that would further restrict the allowed 
parameter space within the 2HDM, briefly outlined in section~\ref{sec:scan2}. 
The reason for not doing so is that these constraints depend significantly on the Type of 2HDM, while our analysis 
of the EWPT and the bounds from stability, unitarity, perturbativity and existence of a panic vacuum do not. However, it is important to briefly
discuss these experimental constraints so that the reader is well informed of their potential impact on the 2HDM parameter 
space:~{\it (i)} LEP searches yield the limit $m_{H^\pm} > 72$ GeV ($80$ GeV) for 2HDM Type I (II)~\cite{Abbiendi:2013hk} as well as the bound 
$m_{H_0} + m_{A_0} \gtrsim 209$ GeV~\cite{Schael:2006cr}.~{\it (ii)} LHC measurements of Higgs signal strengths 
constrain the allowed value of $c_{\beta-\alpha}$ as a function of $t_{\beta}$ 
(see {\it e.g.}~\cite{Dumont:2014wha,Chowdhury:2015yja,Craig:2015jba,Dorsch:2016tab}). These do not provide a constraint in the alignment limit
$c_{\beta-\alpha} = 0$ (since the 125 GeV Higgs behaves as the SM one in this case), but do constrain significant deviations from the 
alignment limit, and thus will be relevant for the analysis of section~\ref{2HDM_NA}. In addition, Higgs signal strength measurements constrain 
the size of the $h \to A_0 A_0$ partial width for $m_{A_0} < 62$ GeV, which in alignment translates into the strong 
constraint $\Omega^2 \simeq m_{H_0}^2 - m_{A_0}^2 - m_h^2/2$ 
on the allowed range of $\Omega$ in this region~\cite{Bernon:2014nxa}.~{\it (iii)} LHC 
searches for $H_0$, $A_0$ and $H^{\pm}$ constrain the masses of the new scalars as a function of $c_{\beta-\alpha}$ and $t_{\beta}$ (and $\Omega$ in 
certain regions of parameter space). In the alignment limit, and for the parameters considered 
in Figs.~\ref{Vacuum_Energy_Alignment_1} and~\ref{Vacuum_Energy_Alignment_2},
relevant limits come from $A_0 \to Z H_0$ ($H_0 \to Z A_0$) 8 TeV CMS searches~\cite{Khachatryan:2016are} 
in the region $\Delta m_{AH} > 0$ ($\Delta m_{AH} < 0$), as discussed 
in~\cite{Dorsch:2016tab}. Searches for $H^{\pm}$ are also relevant for $m_{H^{\pm}} < m_t$ 
(see {\it e.g.}~\cite{Aad:2014kga}).~{\it (iv)} Flavour constraints, particularly from $\bar{B}\to X_s \gamma$ $B$-meson decays, yield strong 
limits on the ($m_{H^{\pm}},\, t_{\beta}$) parameter space both for Type I~\cite{Hermann:2012fc} and 
Type II~\cite{Hermann:2012fc,Misiak:2015xwa} 2HDM (see also~\cite{Misiak:2017bgg}).

\begin{figure}[h!]
\centering

$\vcenter{\hbox{\includegraphics[width=0.385\textwidth]{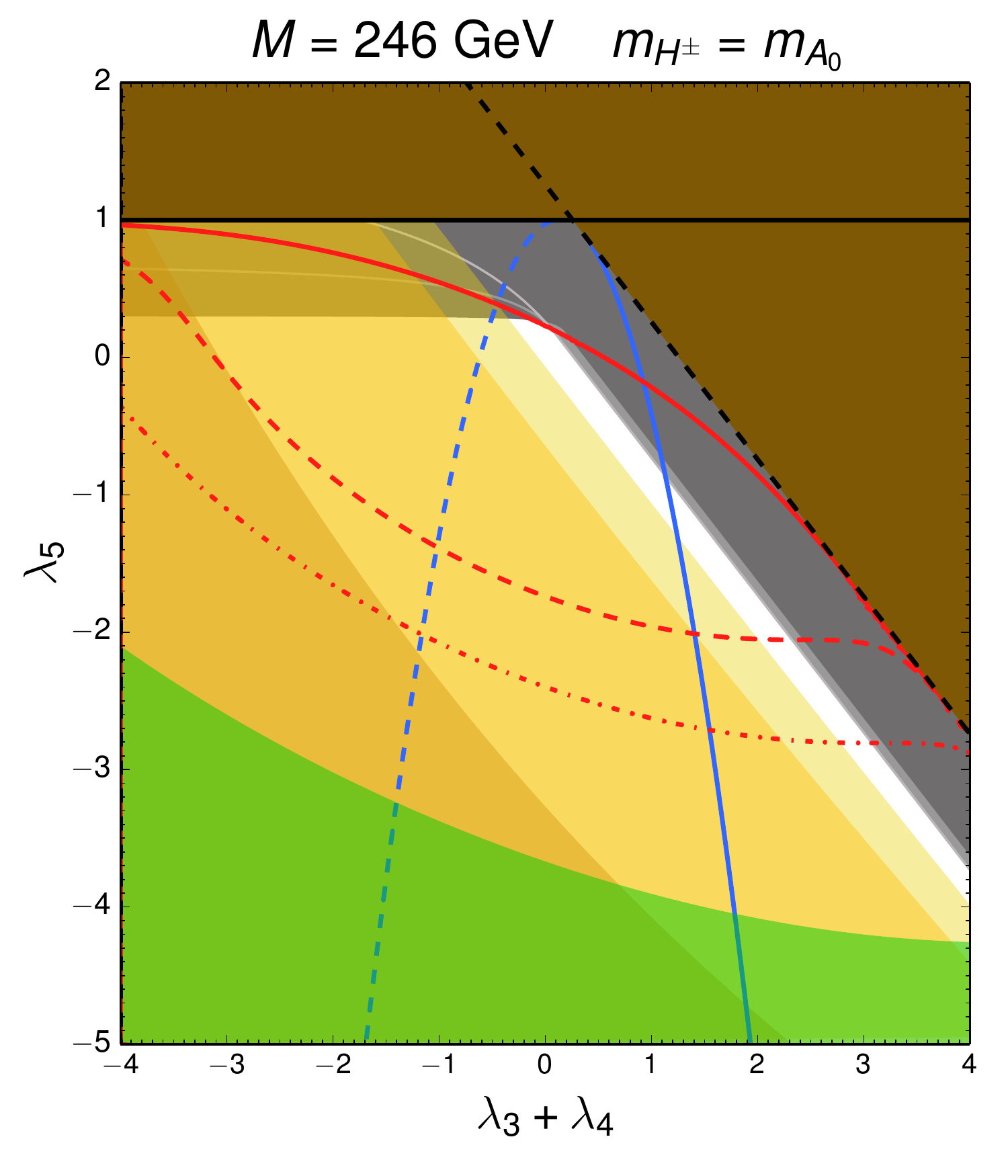}}}$
$\vcenter{\hbox{\includegraphics[width=0.385\textwidth]{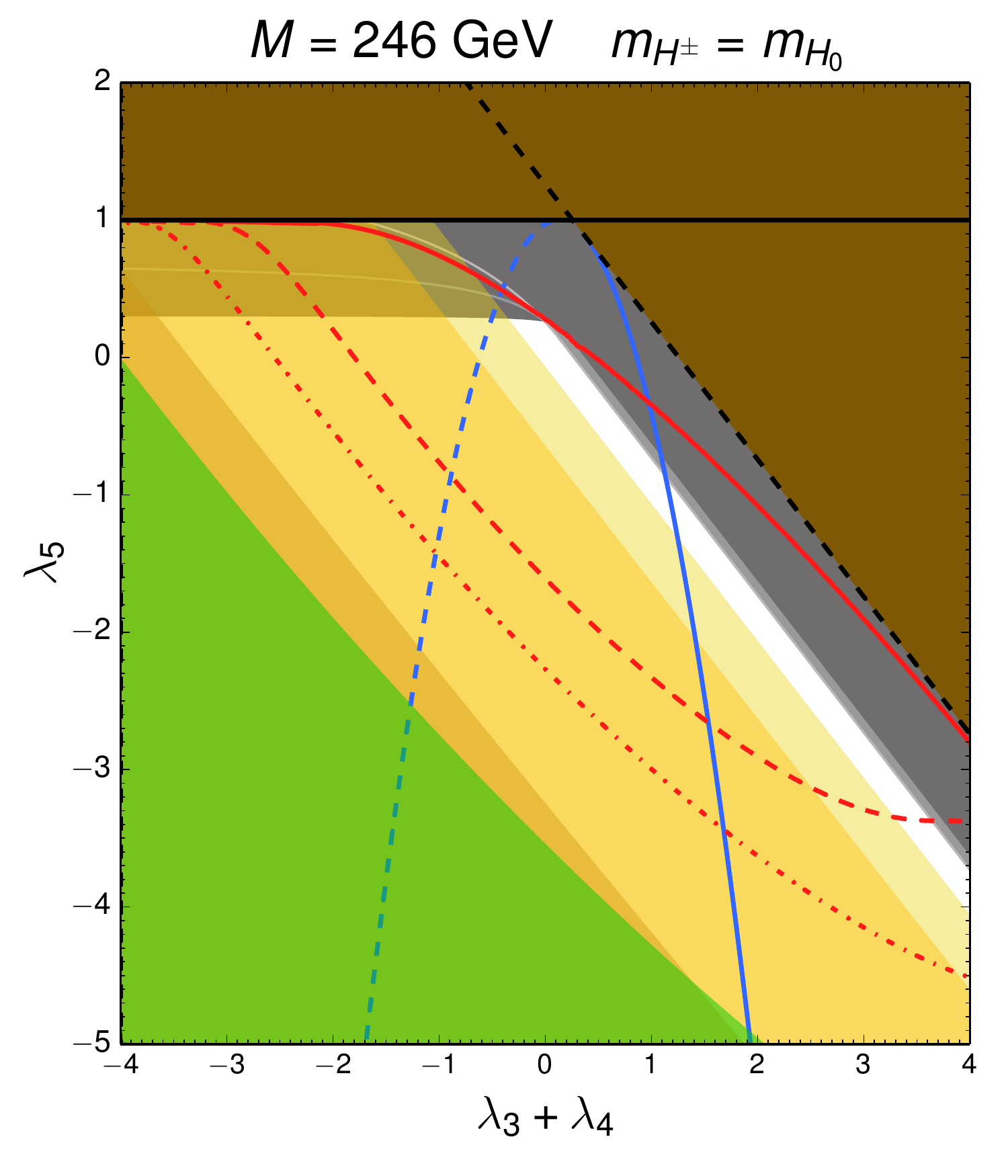}}}$
$\vcenter{\hbox{\includegraphics[width=0.21\textwidth]{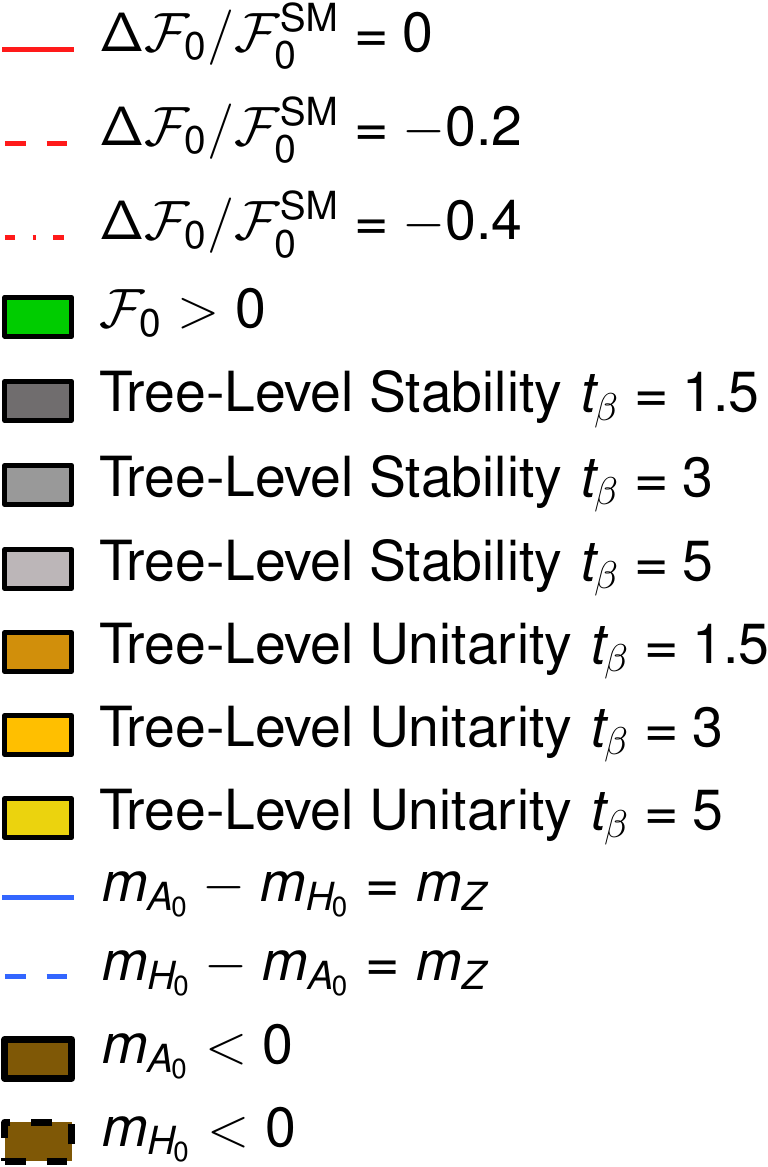}}}$

\caption{\small{$\lambda_3 + \lambda_4$ vs 
$\lambda_5$ for $M = 246$ GeV and assuming respectively $m_{H^{\pm}} = m_{A_0}$ (Left) and 
$m_{H^{\pm}} = m_{H_0}$ (Right). Red lines show constant values of $\Delta\mathcal{F}_0/\mathcal{F}^{\mathrm{SM}}_0$, with 
the green region corresponding to $\Delta\mathcal{F}_0/\mathcal{F}^{\mathrm{SM}} < -1$ ($\mathcal{F}_0 > 0$).
Blue lines show the contours $m_{A_0} - m_{H_0} > m_Z$ (solid) and $m_{H_0} - m_{A_0} > m_Z$ (dashed). 
The grey and orange regions are respectively excluded by boundedness from below of the scalar potential and by unitarity, 
for $t_{\beta} = 1.5,\,3,\,5$ (dark to light). The brown region is unphysical 
($m_{H_0} < 0$ and/or $m_{A_0} < 0$).\label{Vacuum_Energy_Alignment_Quartics}}}
\end{figure}

In order to shed some more light on the impact of the quartic coupling values from the 2HDM potential~\eqref{2HDM_potential} 
on the strength of the EWPT, 
we now analyze the interplay between $\Delta\mathcal{F}_0$ and the theoretical constraints 
using a different choice of independent parameters: $c_{\beta-\alpha}$, $t_{\beta}$, $M^2$, $\lambda_3$, 
$\lambda_4$, $\lambda_5$. Together with $v = 246$~GeV and $m_h = 125$~GeV, these completely determine the parameters in~\eqref{2HDM_potential}.
We fix $c_{\beta-\alpha} = 0$, and note that $\Delta\mathcal{F}_0$ in this limit, given by~\eqref{F0_alignment}, is symmetric under $m_{A_0} \leftrightarrow m_{H_0}$. Fixing $m_{H^{\pm}}$ to be close to either $m_{A_0}$ or $m_{H_0}$ breaks this symmetry.
However, there is still a symmetry between the scenario $m_{H^{\pm}} = m_{A_0}$ with $\Delta m_{AH} > 0$ and the scenario $m_{H^{\pm}} = m_{H_0}$ 
with $\Delta m_{AH} < 0$. Using the relations from Appendix~\ref{app:Z2dict} we find that in the former 
scenario $\lambda_4 = \lambda_5$ while in the latter $\lambda_4 = m_h^2/v^2 - (2\lambda_3 + \lambda_5)$.
In both cases $m^2_{A_0} - m^2_{H_0} = v^2 (\lambda_3 + \lambda_4) - m_h^2$.
Choosing $M = 246$ GeV as an illustrative example, we compare in Fig.~\ref{Vacuum_Energy_Alignment_Quartics} the vacuum energy 
difference $\Delta\mathcal{F}_0$ and theoretical constraints
in the ($\lambda_3 + \lambda_4$, $\lambda_5$) plane, for the $m_{H^{\pm}} = m_{A_0}$ and $m_{H^{\pm}} = m_{H_0}$ scenarios.
In each case, besides the lines of constant $\Delta\mathcal{F}_0/\mathcal{F}_0^{\mathrm{SM}} = 0$, $-0.2$, $-0.4$ and $-1$ ($\mathcal{F}_0 > 0$), 
we show the contours of $m_{A_0} - m_{H_0} = m_Z$ (when the decay $A_0 \to Z H_0$ becomes kinematically accessible) and 
$m_{H_0} - m_{A_0} = m_Z$ (when the decay $H_0 \to Z A_0$ becomes kinematically accessible), as well as the tree-level stability and unitarity bounds for 
$t_{\beta} = 1.5$, $3$, $5$. Fig.~\ref{Vacuum_Energy_Alignment_Quartics} explicitly shows that for $t_{\beta} \sim 1$ sufficient vacuum uplifting 
for a strongly first order EWPT in the 2HDM is compatible with both $m_{A_0} - m_{H_0} > m_Z$ and $m_{H_0} - m_{A_0} > m_Z$ (and even 
$m_{H_0} = m_{A_0}$). This is the case for both the $m_{H^{\pm}} = m_{A_0}$ (Fig.~\ref{Vacuum_Energy_Alignment_Quartics} Left) 
and $m_{H^{\pm}} = m_{H_0}$ (Fig.~\ref{Vacuum_Energy_Alignment_Quartics} Right) scenarios. However, as $t_{\beta}$ increases, the region 
$m_{H_0} > m_{A_0}$ becomes progressively excluded by unitarity, and already for $t_{\beta} = 3$ a vacuum uplifting 
$\Delta\mathcal{F}_0/\mathcal{F}_0^{\mathrm{SM}} = -0.2$ demands $m_{A_0} - m_{H_0} > m_Z$, as can also be inferred from Figs.~\ref{Vacuum_Energy_Alignment_1}
and~\ref{Vacuum_Energy_Alignment_2}.

\begin{figure}[t!]
\centering

$\vcenter{\hbox{\includegraphics[width=0.385\textwidth]{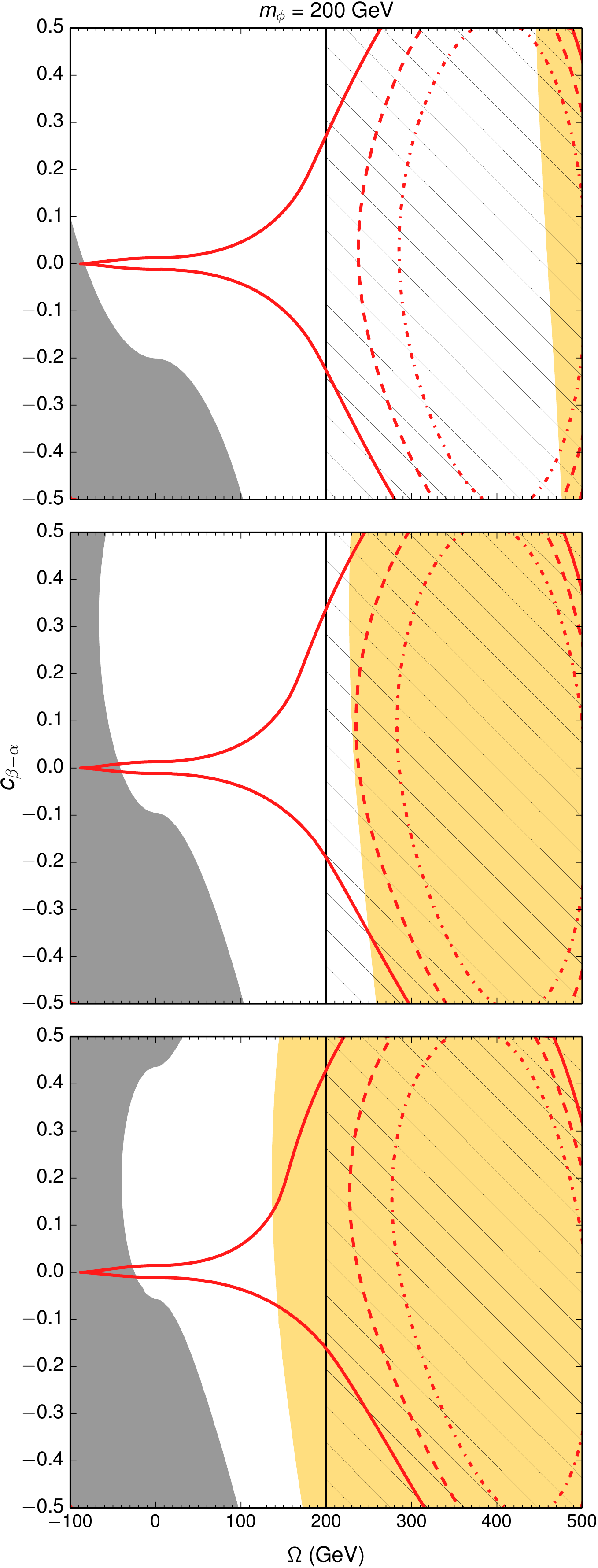}}}$
$\vcenter{\hbox{\includegraphics[width=0.385\textwidth]{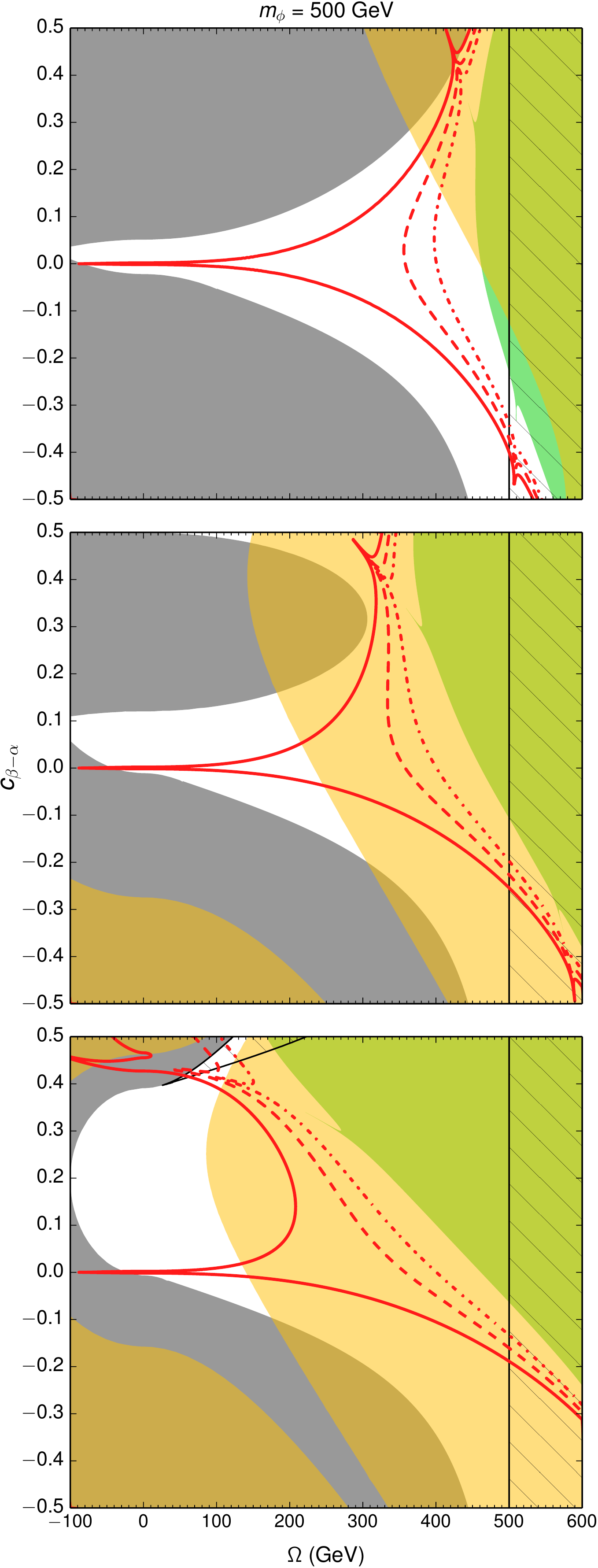}}}$
$\vcenter{\hbox{\includegraphics[width=0.21\textwidth]{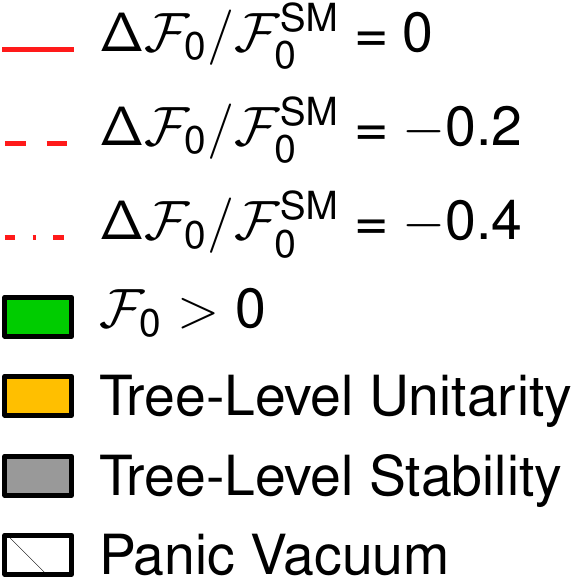}}}$

\caption{\small{$\Omega \equiv \sqrt{\left|\Omega^2\right|}\times\mathrm{sign}(\Omega^2)$ vs 
$c_{\beta-\alpha}$ for 
$m_{\phi} = 200,\,500$ GeV (Left to Right) and $t_{\beta} = 1.5,\,3,\,5$ (Top to Bottom).
Red lines show constant values of $\Delta\mathcal{F}_0/\mathcal{F}^{\mathrm{SM}}_0$. 
The grey region is excluded by boundedness from below of the scalar potential, while the orange region is excluded 
by unitarity. In the hatched region, a panic vacuum develops.\label{Vacuum_Energy_NA}}}
\end{figure}

\subsection{Away from the Alignment Limit: Degenerate 2HDM Spectrum}
\label{2HDM_NA}

We now investigate the effect of departing from the alignment limit, setting for simplicity $m_{H_0} = m_{A_0} = m_{H^{\pm}} = m_{\phi}$. 
In this approximation the vacuum energy difference can be expressed in terms of 
$c_{\beta-\alpha}$,  $t_{\beta}$, $m^2_{\phi}$ and $\Omega^2$ (see Appendix~\ref{app:HB_renorm} for details).
We show in Fig.~\ref{Vacuum_Energy_NA} the behaviour of the vacuum energy difference in the ($\Omega,\,c_{\beta-\alpha}$) plane 
for $m_{\phi} = 200,\,500$ GeV (Left to Right) and $t_{\beta} = 1.5,\,3,\,5$ (Top to Bottom). In all cases a sizable vacuum uplifting demands 
$\Omega \gtrsim v$ (the only exception corresponds to $m_{\phi} = 500$ GeV, $t_{\beta} = 5$ and $c_{\beta-\alpha} \gtrsim 0.4$, excluded by vacuum stability).
As shown in Fig.~\ref{Vacuum_Energy_NA} (Left), for light $m_{\phi}$ uplifting of the vacuum is in conflict with the panic vacuum constraint (and also 
excluded by unitarity for $t_{\beta} \gg 1$). In contrast, Fig.~\ref{Vacuum_Energy_NA} (Right) shows that sufficient vacuum uplifting is possible 
for $m_{\phi} = 500$ GeV and $v \lesssim \Omega \lesssim m_{H_0}$, provided that $t_{\beta} \sim 1$. Again, as $t_{\beta}$ increases the parameter space 
region where the 2HDM Higgs vacuum is uplifted compared to the SM one becomes excluded by unitarity.

\subsection{An Inert Second Doublet\label{sec:I2HDM}}

The inert doublet model~\cite{Ma:2006km,Barbieri:2006dq,LopezHonorez:2006gr} (IDM) 
is a special case of 2HDM scenario in which the second doublet is protected by a $\mathbb{Z}_2$ symmetry and does not develop a vev.
This $\mathbb{Z}_2$ symmetry leads to the lightest state of the second doublet being stable, yielding a viable dark matter (DM) candidate 
if this corresponds to either $A_0$ or $H_0$. This scenario has been widely studied in the literature (see e.g.~\cite{Arhrib:2013ela,Ilnicka:2015jba} 
for updated analyses, and references therein), including its impact on the EWPT~\cite{Chowdhury:2011ga,Borah:2012pu,Gil:2012ya,Blinov:2015vma}.

The scalar potential for the IDM is given by~\eqref{2HDM_potential} with $\mu = 0$, and due to the unbroken $\mathbb{Z}_2$ symmetry
the dictionaries from Appendix \ref{app:Z2dict} -- \ref{app:HBdict} do not apply in any particular limit, and instead the relations among parameters 
are given in \ref{app:IDMdict} (note however that some of the parameter relations are identical to those of the Higgs basis 
with $c_{\beta-\alpha} = 0$ and $M^2 = 0$). The relevant IDM parameters can be conveniently chosen to be $m_{H_0}$, $m_{A_0}$, $m_{H^{\pm}}$, 
$\lambda_{345} \equiv \lambda_3 + \lambda_4 + \lambda_5$ and $\lambda_2$. 
In the following we consider DM to be $H_0$ (both choices are physically equivalent in the IDM), 
which amounts to requiring $\Delta m_{AH} > 0$, and we also consider $m_{H^{\pm}} = m_{A_0}$ as a simplifying assumption to satisfy EW precision constraints.

Using~\eqref{2HDM_counterterms},~\eqref{2HDM_potential5} and the results from Appendix~\ref{app:HB_renorm} we can easily obtain the vacuum energy 
difference $\Delta\mathcal{F}_0$ for the IDM, which reads
\begin{eqnarray}
\label{F0_IDM}
	\Delta\mathcal{F}_0 &=& \frac{1}{64\,\pi^2} \left[ \left(m_{H_0}^2- \frac{\lambda_{345}v^2}{2} \right)^2 
	\, \mathrm{log} \left[ \frac{m_{H_0}^2\,m_{A_0}^6}{\left(m_{H_0}^2- \frac{\lambda_{345}v^2}{2} \right)^4 } \right]
	  + \frac{1}{2} \, (m_{A_0}^4-m_{H_0}^4)  + 3 \left(\frac{\lambda_{345}v^2}{2} \right)^2\right. \nonumber \\
	  & & \left. +\, 4\, \left(m_{H_0}^2-m_{A_0}^2- \frac{\lambda_{345}v^2}{2} \right)  \left(m_{H_0}^2- \frac{\lambda_{345}v^2}{2} \right)
	  + \left(m_{H_0}^2-m_{A_0}^2 - \frac{\lambda_{345}v^2}{2} \right)^2  \right] \, ,
\end{eqnarray}
and we investigate its interplay with theoretical constraints: stability, unitarity and the requirement that the $\mathbb{Z}_2$ symmetry is preserved in 
the EW broken vacuum, which leads to the condition
\begin{equation}
\label{Condition_IDM}
\mu^2_1/\sqrt{\lambda_1} < \mu^2_2/\sqrt{\lambda_2} \, .
\end{equation}
We also include in our analysis the constraint on the IDM parameter space from the latest LUX bounds on the spin-independent DM-nucleon 
scattering cross section~\cite{Akerib:2016vxi}, as well as the IDM parameter space region for which the $H_0$ relic abundance through thermal freeze-out 
$\Omega_{H_0}$ does not exceed the observed 
DM relic density $\Omega_{\mathrm{DM}} = 0.1199\pm 0.0022$~\cite{Ade:2015xua}. The $H_0$ relic abundance and the 
spin-independent $H_0$-nucleon scattering cross section are both obtained with {\sc micrOMEGAs$\_$4.3}~\cite{Belanger:2013oya}, and we note that 
the nucleon scattering cross section has to be weighted by $\Omega_{H_0}/\Omega_{\mathrm{DM}}$ when comparing with the LUX limits 
(as these assume $\Omega_{H_0} = \Omega_{\mathrm{DM}}$).

\begin{figure}[h!]
\centering

$\vcenter{\hbox{\includegraphics[width=0.385\textwidth]{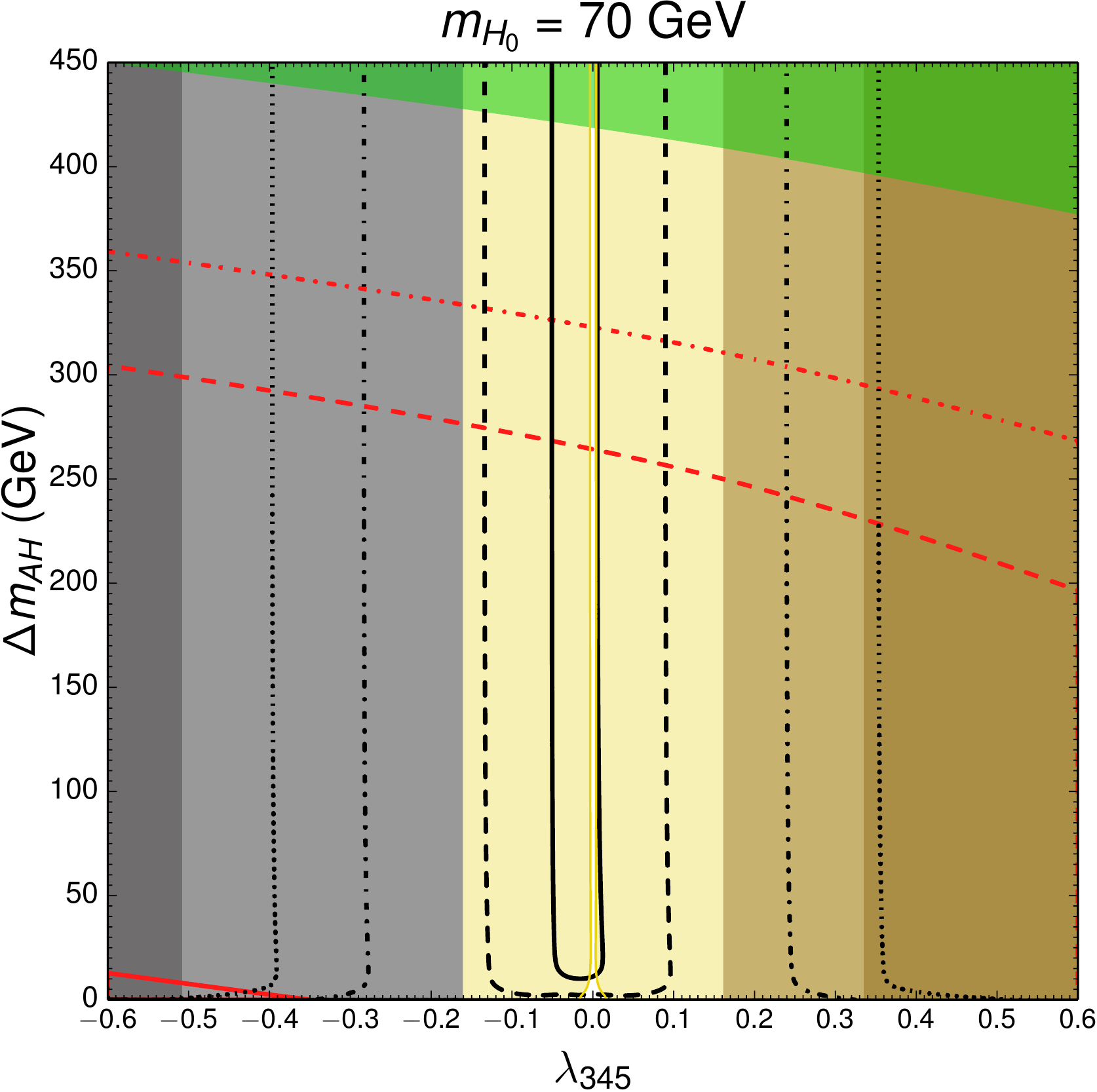}}}$
$\vcenter{\hbox{\includegraphics[width=0.385\textwidth]{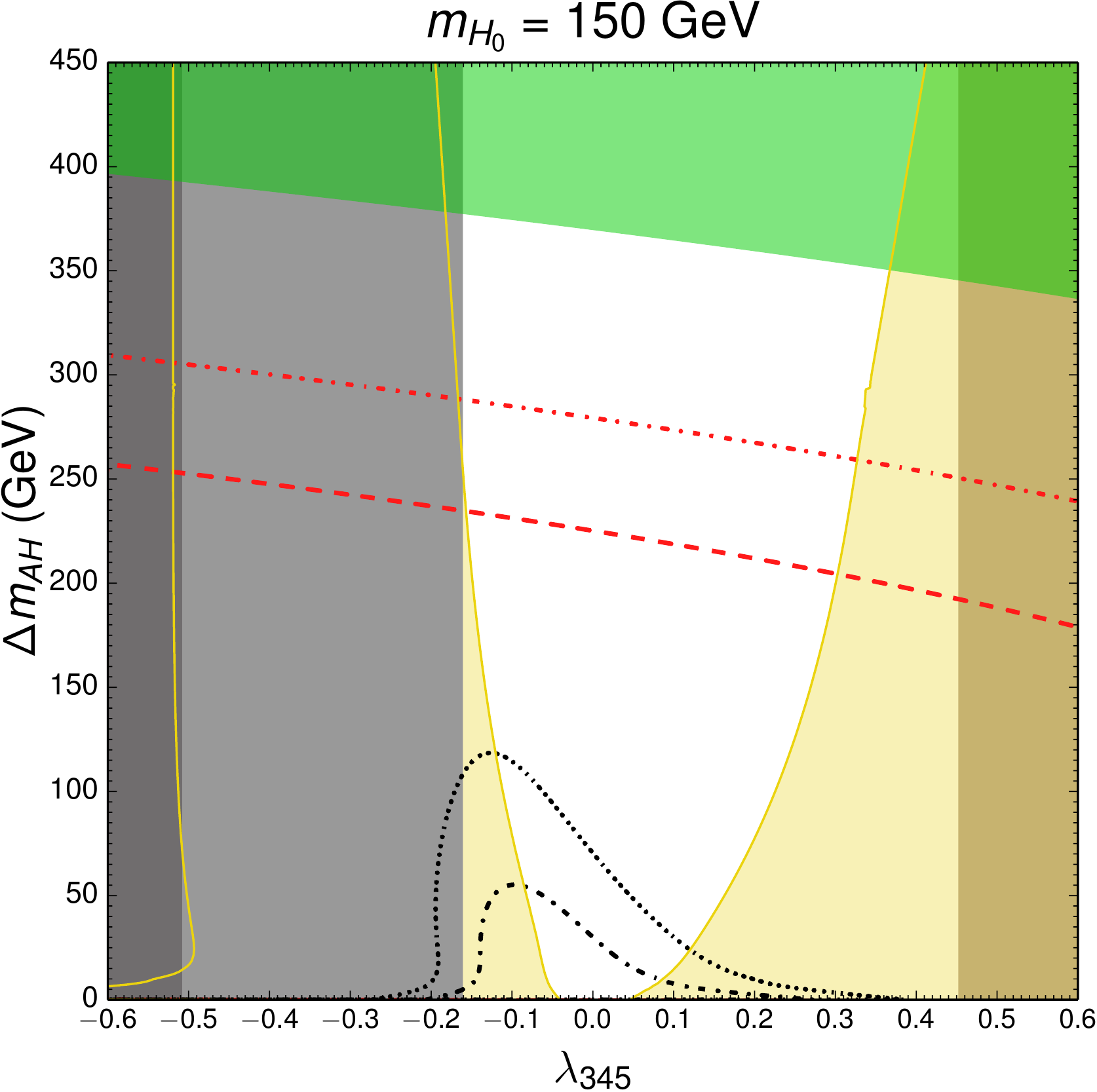}}}$
$\vcenter{\hbox{\includegraphics[width=0.21\textwidth]{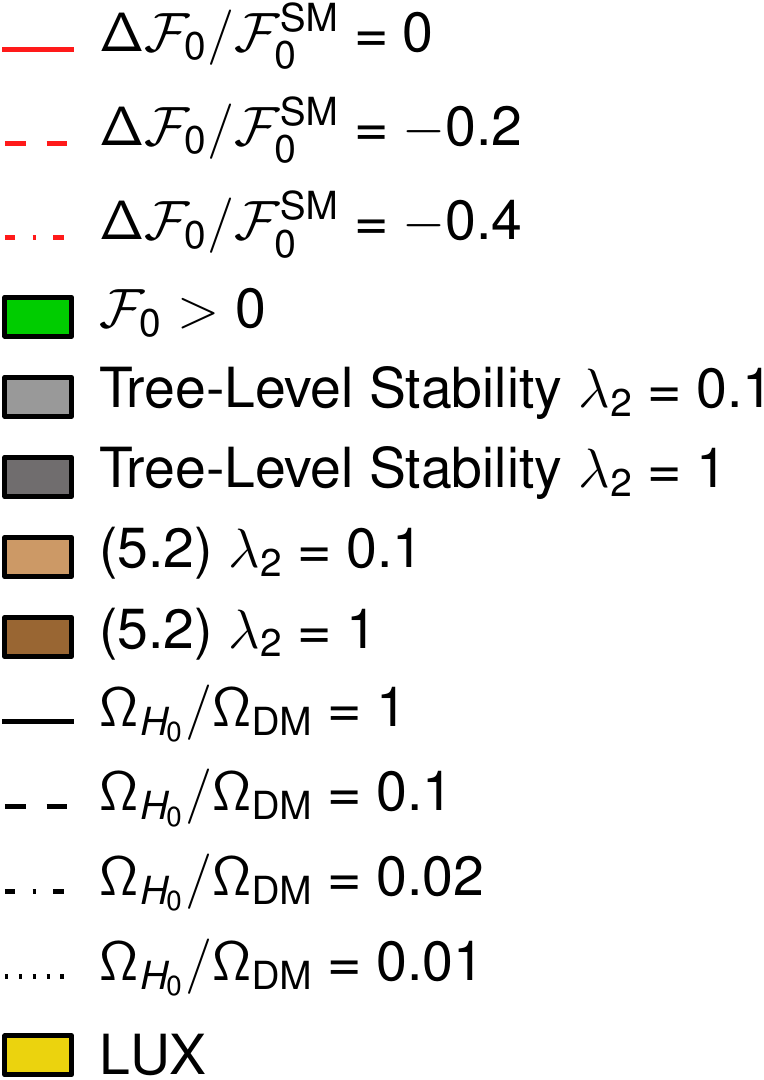}}}$

\caption{\small{$\lambda_{345}$ vs 
$\Delta m_{AH}$ assuming $m_{H^{\pm}} = m_{A_0}$, for $m_{H_0} = 70$ GeV.
Red lines show constant values of $\Delta\mathcal{F}_0/\mathcal{F}^{\mathrm{SM}}_0$, with 
the green region corresponding to $\Delta\mathcal{F}_0/\mathcal{F}^{\mathrm{SM}} < -1$ ($\mathcal{F}_0 > 0$).
The grey and brown regions are respectively excluded by boundedness from below of the scalar potential and by the failure to 
fulfill eq.~\eqref{Condition_IDM}, respectively for $\lambda_2 = 1$ (dark) and $\lambda_2 = 0.1$ (light). 
Contours of constant $\Omega_{H_0}/\Omega_{\mathrm{DM}} = 1, \,0.1,\,0.02,\, 0.01$  are shown as solid, dashed, dash-dotted and dotted 
black lines. The excluded region from LUX~\cite{Akerib:2016vxi} is shown in pale yellow. \label{fig:IDM}}}
\end{figure}

In Fig.~\ref{fig:IDM} we show the vacuum energy difference in the plane ($\lambda_{345},\,\Delta m_{AH}$) for benchmark values
$m_{H_0} = 70$ GeV (left) and $m_{H_0} = 150$ GeV (right), as well as the theoretical constraints for $\lambda_2 = 1,\,0.1$. We also show 
the contours of constant $\Omega_{H_0}/\Omega_{\mathrm{DM}} = 1, \,0.1,\, 0.02,\,0.01$ and the bound from LUX.
For $m_{H_0} = 70$ GeV the LUX bound combined with $\Omega_{H_0}/\Omega_{\mathrm{DM}} \leq 1$ exclude 
the entire parameter space except for the small island $\Delta m_{AH} \lesssim 10$ GeV and $  -0.05 \lesssim \lambda_{345} \lesssim 0.05$.
As shown in Fig.~\ref{fig:IDM} significant vacuum uplifting requires $\Delta m_{AH} \gtrsim v$ and is thus not possible in this 
case\footnote{We note that for this value of $m_{H_0}$ a strong EWPT was deemed possible in~\cite{Blinov:2015vma}, but we find 
the most recent LUX limits exclude this possibility.}. 
In  contrast for $m_{H_0} = 150$ GeV, sizable uplifting and thus a strongly first order EWPT is possible, requiring $\Delta m_{AH} \gtrsim 200$ GeV.
We emphasize that while previous works have already identified a large mass splitting $\Delta m_{AH}$ 
in the IDM as providing a strong EWPT (see e.g.~\cite{Blinov:2015vma}), the dominant strengthening effect was attributed to the thermal 
contributions of $H_0$, $A_0$, $H^{\pm}$ to $V^{T}_{\mathrm{eff}}$. While these do play an important role, 
we show here that the most important effect is due to the uplifting of the $T=0$ vacuum.

\section{Trilinear Higgs self-coupling\label{sec:trilinear}}

\begin{figure}[t]
\centering
\includegraphics[width=\textwidth]{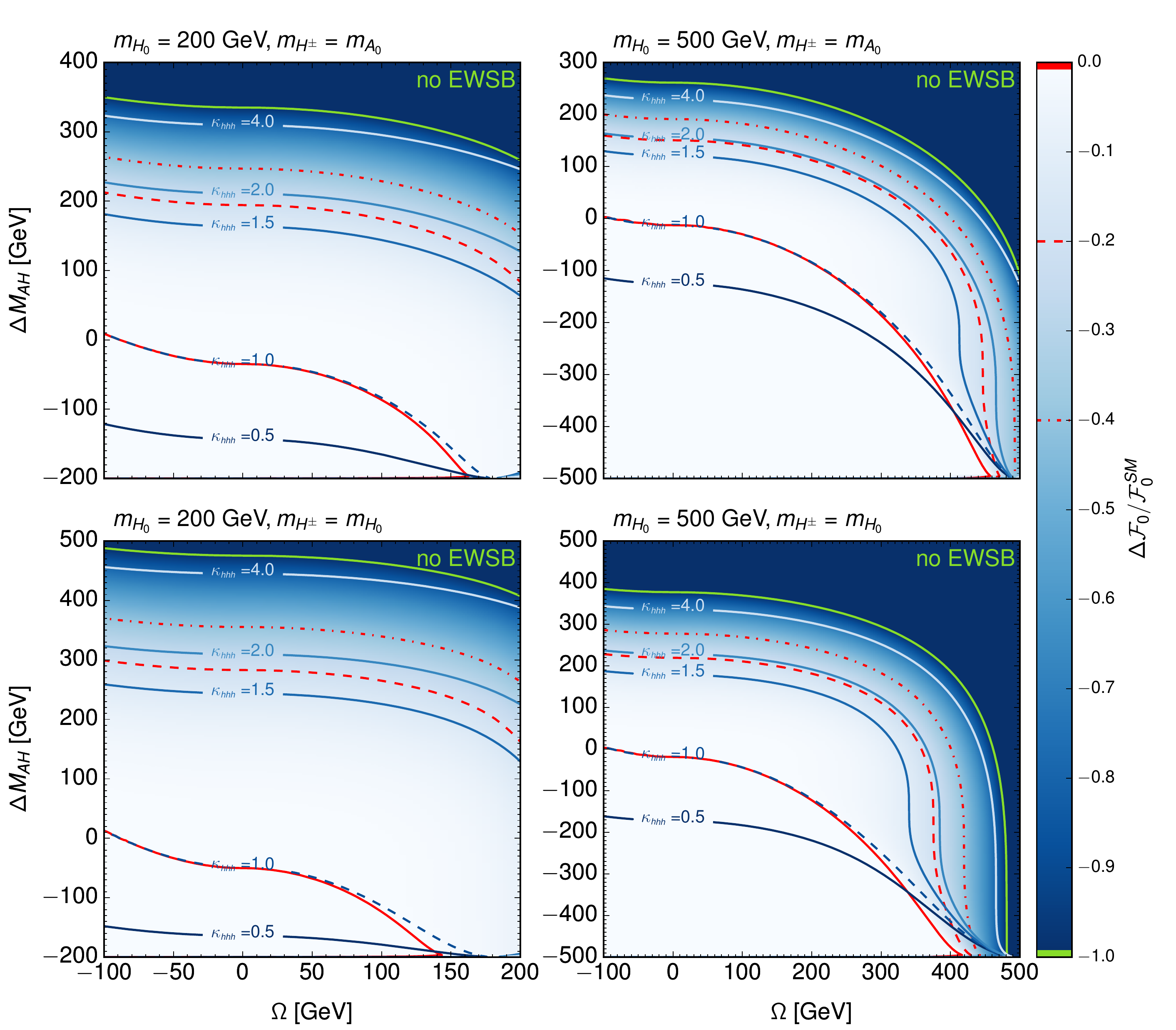}
\caption{\small Contours of the deviation in the 2HDM Higgs self-coupling $\kappa_{hhh}=\lambda_{hhh}/\lambda^{\mathrm{SM}}_{hhh}$ 
overlaying the vacuum energy difference. 
The dashed curve corresponds to $\kappa_{hhh}= 1$, where the prediction is unchanged with respect to the SM. The  
values of 1.5 and 0.5 correspond to the expected precision envisaged for the HL-LHC. Vacuum energy difference values of 0 and -1 are also 
highlighted in which either no EWSB can occur or the vacuum energy difference is the same as in the SM respectively.
\label{trilinear}}
\end{figure}

Finally, it is useful to discuss the behaviour of the trilinear Higgs self-coupling $\lambda_{hhh}$ in the  
($\Omega$, $\Delta m_{\mathrm{AH}}$) plane, w.r.t. its value in the SM $\lambda^{\mathrm{SM}}_{hhh}$. 
It has been suggested that a strong first order EWPT in the 2HDM is tightly correlated with 
sizable deviation in the value of $\lambda_{hhh}$ w.r.t. the SM value~\cite{Kanemura:2002vm,Kanemura:2004ch}.
In the alignment limit, we note that $\lambda_{hhh} = \lambda^{\mathrm{SM}}_{hhh}$ at tree-level (as was also noted 
in~\cite{Kanemura:2002vm,Kanemura:2004ch}). However, in the 2HDM 1-loop corrections may lead to sizable deviations from the SM value.
The Higgs self-coupling $\lambda_{hhh}$ in the 2HDM is approximately given at 1-loop 
by
\begin{align}\nonumber
        \lambda_{hhh}=&\frac{3m_h^2}{v} + \sum_{k}
 \frac{n_{k}}{32\pi^2v^3}\frac{(vI_k)^3}{m_k^2}\\\nonumber
 =&\frac{3 m_h^2}{v} \left(1 - \frac{m_t^4}{\pi^2\,m_h^2\,v^2} \right)+\sum_{k=H_0,A_0,H^\pm}n_{k}\,\frac{m_k^4}{4\pi^2v^3}
 \left(1+\frac{m_h^2}{2m_k^2} - \frac{M^2}{m_k^2}\right)^3\\ =&\lambda_{hhh}^{\mathrm{SM}}+\sum_{k=H_0,A_0,H^\pm}n_{k}\,\frac{m_k^4}{4\pi^2v^3}
 \left(1+\frac{m_h^2}{2m_k^2} - \frac{M^2}{m_k^2}\right)^3
\end{align}
where $\lambda^{\mathrm{SM}}_{hhh}$ includes the SM 1-loop corrections due to the top quark, Higgs and gauge bosons. 
Our result agrees with~\cite{Kanemura:2002vm,Kanemura:2004ch} and includes some sub-leading pieces that become relevant when the new 
scalar states are not so heavy with respect to the 125 GeV Higgs boson.
Given the tight correlation between the vacuum energy difference and the strength of the EWPT, one would also expect 
a relationship to exist between the former and the Higgs self-coupling. Defining $\kappa_{hhh}\equiv\lambda_{hhh}/\lambda^{\mathrm{SM}}_{hhh}$, 
the region $\left|1-\kappa_{hhh}\right| \geq 0.5$ is of particular interest, since such a deviation in $\lambda_{hhh}$ from its SM value 
could be probed at the HL-LHC~\cite{Yao:2013ika, Dawson:2013bba}. In Fig.~\ref{trilinear} we show contours of $\kappa_{hhh}$, 
for $m_{H_0} = 200$ GeV and $m_{H_0} = 500$ GeV in both $m_{H^\pm}=m_{H_0}$ and $m_{H^\pm}=m_{A_0}$ scenarios. We 
also superimpose the normalized vacuum energy difference $\Delta\mathcal{F}_0/\mathcal{F}^{\mathrm{SM}}_0$, highlighting (in red/green) the 
values 0 and -1. The latter 
case corresponds to the limit above which the EW vacuum is lifted above the trivial one ($\mathcal{F}_0 > 0$), preventing EWSB from ever occurring, 
while the former denotes a vacuum energy difference equal to that of the SM. Interestingly, we see that the region of unchanged vacuum 
energy difference with respect to the SM coincides almost exactly with the region where the Higgs self-coupling does not deviate from the 
SM prediction. Furthermore, the self-coupling grows as the EW vacuum is uplifted, reaching values of 2-4 times the SM prediction in the regions 
shown in Figs.~\ref{Vacuum_Energy_Alignment_1} and~\ref{Vacuum_Energy_Alignment_2} where a strong EWPT is expected to occur.

The strong correlation between the vacuum energy and the trilinear Higgs coupling shown in 
Fig.~\ref{trilinear} can qualitatively be understood in terms of an effective potential for the SM Higgs. 
The extra Higgs states induce higher dimensional operators, with the leading one being of mass dimension six. When 
only keeping the mass term, the quartic coupling and the dimension-6 operator in the Higgs potential, we can vary the 
vacuum energy independently of the Higgs mass and trade the coefficient of the dimension-6 operator for the vacuum energy to 
parametrize this effective potential~\cite{Harman:2015gif}. We can then compute the third derivative of this potential 
to obtain the trilinear Higgs coupling. Setting this in ratio to the SM result, which corresponds to a vanishing dimension-6 operator, we obtain
\begin{align}
\kappa_{hhh}=\frac{3m_h^2v^2+16\mathcal{F}_0}{3m_h^2v^2+16\mathcal{F}_0^\text{SM}}.
\end{align}
Clearly $\mathcal{F}_0>\mathcal{F}_0^\text{SM}$ means $\kappa_{hhh}>1$. Quantitatively, we find that this estimate falls short of the full result in 
Fig.~\ref{trilinear} up to about 30\%. This is not surprising, as the Higgs states integrated out are not very much heavier 
than $v$. So we expect operators of higher mass dimension to play a role, which, however, do not spoil the overall qualitative  picture. 

In fact, the contribution to the dimension-6 operator affecting the Higgs potential from integrating out the new states in the 2HDM is known~\cite{Gorbahn:2015gxa}. Only one operator 
\begin{align}
    \mathcal{O}_6 = \lambda\frac{\bar{c}_6}{v^2}\left(\Phi^\dagger\Phi\right)^3
\end{align}
plays a role here. Its effect of the vacuum energy difference and the Higgs trilinear coupling is as follows
\begin{align}
    \kappa_{hhh} = 1+\bar{c}_6,\quad
    \frac{\Delta\mathcal{F}_0}{|\mathcal{F}_0^\text{SM}|}=1+\frac{\bar{c}_6}{2}.
\end{align}

In the alignment limit, the Wilson coefficient of interest has been calculated as
\begin{align}
    \bar{c}_{6} &=(\bar{\lambda}_{4}^2+\bar{\lambda}_{5}^2)\frac{v^2}{192\pi^2\bar{\mu}_{2}^2}\\
    &= \frac{(m_{A_0}^2 - m_{H^\pm}^2)^2+(m_{H_0}^2 - m_{H^\pm}^2)^2}
                        {48\pi^2v^2(2M^2-m_h^2)}.
\end{align}
Being positive definite, we see that it contributes both to an uplifting of the EW vacuum and an increase in the Higgs trilinear coupling. Furthermore, since EW precision tests constrain the charged Higgs mass to be near one or the other neutral state ($m_{H^\pm}\sim_{H_0}$ or $m_{H^\pm}\sim_{A_0}$), we are left with precisely the aforementioned mass splitting between the two, new neutral states controlling the effects of interest, lending further support to our previous findings.

\section{Conclusions\label{sec:conclusions}}
	
In this work we have established a correlation between the strength of the electroweak phase transition and the zero-temperature free-energy of the broken minimum in two-Higgs-doublet models. Considering similar statements made previously in the literature in the context of other SM scalar sector extensions~\cite{Huang:2014ifa,Harman:2015gif}, we claim this is a general effect of any model where the modified scalar sector acts as the main source of strong phase transition.

Because working with the zero-temperature vacuum energy is analytically much simpler than with the full thermal potential, this correlation can be used to better predict the behaviour of a certain model concerning the nature of the EWPT, as well as to better understand the impact of parameter space constraints on the strength of the phase transition predicted by the model. In particular, we have in this way clarified the preferred hierarchy in the scalar sector from the requirement of a strong EWPT, with a heavier pseudoscalar and charged scalar.

We have further investigated the relation between the triple Higgs self-coupling and the vacuum energy uplifting in the model. Large deviations from the SM predictions of these couplings are expected as a collateral effect of a model with a strong EWPT, and we have shown that these deviations can be measurable at the HL-LHC in some scenarios here presented. A measurement of the Higgs self-couplings is a key goal in any future collider experiment as a probe of the ultimate structure of the Higgs potential. Results such as the ones we present here show that this measurement would also serve as an indirect probe for the nature of the nature of the electroweak phase transition, and of the viability of electroweak baryogenesis as an explanation for the baryon asymmetry of the Universe.


\section*{Acknowledgment}
The work of S.H and K.M. is supported by the Science Technology and Facilities Council (STFC) under grant number
ST/L000504/1. K. M. is also supported in part by the Belgian Federal Science Policy Office through the Interuniversity Attraction Pole P7/37. J.M.N. is supported by the European Research Council under the
European Union’s Horizon 2020 program (ERC Grant Agreement no.648680 DARKHORIZONS). 
G.C.D. is supported by the German Science Foundation (DFG) under the Collaborative Research Center (SFB) 676 Particles, Strings and
the Early Universe.

\appendix

\section{Physical dictionaries of the $\mathbb{Z}_2$ and Higgs bases for two Higgs doublets}
\label{app:dictionary}
Here we provide the detailed expressions for the scalar potential parameters of the 2HDM as a function of the masses and mixings of the scalar sector. 
We define $\Omega^2\equiv m^2_{H_0}-\mu^2(t_\beta+t_\beta^{-1})$.
\subsection{$\mathbb{Z}_2$ basis}
\label{app:Z2dict}
See eq.~\eqref{2HDM_potential} for the definition of the potential parameters.
\begin{eqnarray} 
\begin{split} \label{eqn:Z2dict}
\mu_1^2 & =  \mu^2 t_\beta - \frac{1}{2}\left[
m_{h}^2 + (m_{H_0}^2-m_{h}^2)
c_{\beta-\alpha}\left(c_{\beta-\alpha}+s_{\beta-\alpha}t_\beta \right)
\right],
\\
\mu_2^2 & = \mu^2 t_\beta^{-1} - \frac{1}{2}\left[
m_{h}^2 + (m_{H_0}^2-m_{h}^2)
c_{\beta-\alpha}\left(c_{\beta-\alpha}-s_{\beta-\alpha}t_\beta^{-1}\right)
\right]
\end{split}
\end{eqnarray}
\begin{eqnarray}
\begin{split}\label{eq:lambda_parameters}
v^2 \lambda_1 &=  m_h^2  + \Omega^2 t_\beta^2-
(m_{H_0}^2-m_h^2)
\left[1-
(s_{\beta-\alpha}+c_{\beta-\alpha}t_\beta)^2
\right]t_\beta^2, \\
v^2 \lambda_2 &=  m_h^2  + \Omega^2 t_\beta^{-2}-
(m_{H_0}^2-m_h^2)
\left[
1-(s_{\beta-\alpha}-c_{\beta-\alpha}t_\beta^{-1})^2
\right]t_\beta^{-2},\\
v^2 \lambda_3 &=  2 m_{H^{\pm}}^2 + \Omega^2 - m_h^2
-(m_{H_0}^2-m_h^2)\left[1+
(s_{\beta-\alpha}+c_{\beta-\alpha}t_\beta^{-1})
(s_{\beta-\alpha}-c_{\beta-\alpha}t_\beta)
\right],\\
v^2 \lambda_4 &= m_{A_0}^2-  2 m_{H^{\pm}}^2 + m_{H_0}^2 - \Omega^2 \,,\nonumber\\
v^2 \lambda_5 &=  m_{H_0}^2 - m_{A_0}^2 - \Omega^2   \,.
\end{split}
\end{eqnarray}

\subsection{Higgs basis}
\label{app:HBdict}
See eq.~\eqref{2HDM_potential2} for the definition of the potential parameters.

\begin{eqnarray} 
\begin{split}
\label{modifiedmasses} 
\bar{\mu}_1^2 & =  -\frac{1}{2}\left[
    m_{h}^2 + (m_{H_0}^2-m_{h}^2) c^2_{\beta-\alpha} 
    \right] < 0 \\
\bar{\mu}_2^2 & = -\Omega^2 +\frac{1}{2}
    m_{h}^2 + \frac{1}{2}(m_{H_0}^2-m_{h}^2) \Big[1+
    s_{\beta-\alpha} \left(
    s_{\beta-\alpha} - c_{\beta-\alpha}(t_\beta-t_\beta^{-1})
    \right)
    \Big]
\\
\bar{\mu}^2 & =  - (m_{H_0}^2-m_{h}^2) s_{\beta-\alpha} c_{\beta-\alpha} 
\end{split}
\end{eqnarray}

\begin{equation}
\begin{split}
\label{modifiedcouplings}
&v^2\bar{\lambda}_1= -2 \bar{\mu}_1^2    \\
&v^2\bar{\lambda}_2= m_h^2 + \Omega^2 (t_\beta-t_\beta^{-1})^2
+(m_{H_0}^2-m_{h}^2)
\left[1-(s_{\beta-\alpha} - c_{\beta-\alpha}(t_\beta-t_\beta^{-1}))^2\right]\\
&v^2\bar{\lambda}_3=2m_{H^\pm}^2 -2\bar{\mu}^2_2
\\
&v^2\bar{\lambda}_4=m_{A^0}^2-2 m_{H^\pm}^2+ m_{h}^2
+(m_{H^0}^2-m_{h}^2) s^2_{\beta-\alpha}   \\
&v^2\bar{\lambda}_5= - m_{A^0}^2 +m_{h}^2
+(m_{H^0}^2-m_{h}^2) s^2_{\beta-\alpha}\\
&v^2\bar{\lambda}_6= 2\bar{\mu}^2\\
&v^2\bar{\lambda}_7=-\Omega^2 (t_\beta-t_\beta^{-1})
-(m_{H_0}^2-m_{h}^2)
c_{\beta-\alpha}\left(s_{\beta-\alpha} - c_{\beta-\alpha}(t_\beta-t_\beta^{-1})\right)\\
\end{split}
\end{equation}
The Higgs basis does allow to read in a straightforward way the masses for the new scalars in the symmetric and broken EW phases, which is what will enter into 
the vacuum energy difference.

\subsection{Inert Doublet Model}
\label{app:IDMdict}
The potential parameters in this case are defined by eq.~\eqref{2HDM_potential}, with $\mu^2 = 0$.
\begin{eqnarray} 
\begin{split}
\label{IDMparameters} 
\mu_1^2 & =  -\frac{m_{h}^2}{2} \\
\mu_2^2 & = m_{H_0}^2 - \frac{\lambda_{345}}{2}\,v^2 \\
v^2\lambda_1 & = m_{h}^2\\
v^2\lambda_3 & = 2 \left(m_{H^{\pm}}^2  - m_{H_0}^2 \right) + \lambda_{345} \,v^2\\
v^2\lambda_4 & = m_{H_0}^2 + m_{A_0}^2 - 2 \, m_{H^{\pm}}^2 \\
v^2\lambda_5 & = m_{H_{0}}^2  - m_{A_0}^2 
\end{split}
\end{eqnarray}
with $\lambda_{345} \equiv \lambda_3 + \lambda_4 + \lambda_5$, $\lambda_2$ and the scalar masses $m_{H_0}$, $m_{A_0}$, $m_{H^\pm}$ as independent parameters.

\section{On-Shell Renormalization of the 2HDM: $\mathcal{F}_0$ in the Higgs basis}\label{app:HB_renorm}
We recall the scalar contribution to the zero-temperature 2HDM vacuum energy in the basis of~\ref{2HDM_potential2} (eq.~\eqref{F_0})
\begin{eqnarray}	
\begin{split}\label{eqn:F_0_APP}  
\mathcal{F}_0 =& -\frac{m_h^2 v^2}{8} - \frac{v^2}{8}c^2_{\beta-\alpha}\,(m_{H_0}^2 - m_h^2) + \Delta V_1
- \frac{\delta \bar{\mu}_1^2 \,v^2}{2} 
+ \frac{\delta \bar{\lambda}_1 \,v^4}{8}.
\end{split}
\end{eqnarray}
The first two terms correspond to the tree-level piece, $-\bar{\lambda}_1 v^4/8$, translated with eq.~\eqref{modifiedmasses}. The second half of the expression is the 1-loop correction, comprising of the difference between the Coleman Weinberg potential evaluated at the EW minimum and the origin as well as the relevant counterterms. The latter are chosen to preserve the tree-level minimum and scalar masses at 1-loop, which fixes their value to
\begin{align}\label{eqn:counter_terms}
    \delta \bar{\mu}_1^2 
    \equiv
    \frac{1}{2}\left(
    \left.\frac{\partial^2 V_1}{\partial h_1^2} \right|_v- \left.\frac{3}{v}\frac{\partial V_1}{\partial h_1}\right|_v 
    \right),\quad
    \delta \bar{\lambda}_1 
    \equiv
    \frac{1}{v^2}\left(
    \left.\frac{\partial^2 V_1}{\partial h_1^2} \right|_v- \left.\frac{1}{v}\frac{\partial V_1}{\partial h_1}\right|_v 
    \right),
\end{align}
with\footnote{Note that there is a caveat in carrying out the condition in eq.~(\ref{eqn:d2V1}). 
For the Goldstone bosons, the first term in eq.~(\ref{eqn:d2V1}) is infrared divergent, so that trying to define the physical mass by taking derivatives 
of $V_{\rm eff}$ actually yields unphysical results. This happens because, by definition, the effective potential takes into account only diagrams with 
vanishing external momenta, whereas the physical mass must be evaluated on-shell, with $p^2=m^2$. A rigorous solution to the problem has been developed 
in~\cite{Cline:1996mga}, and also in~\cite{Martin:2014bca,Elias-Miro:2014pca} via resummation of the Goldstone contributions. Here we choose 
to adopt the more straightforward approach of replacing the vanishing Goldstone masses in the logarithmic divergent term by an IR cutoff at $m_{\rm IR}^2=m_{h^0}^2$, 
which gives a good approximation to the exact procedure of on-shell renormalization, as argued in~\cite{Cline:2011mm}.}
\begin{align}
\label{eqn:dV1}
\frac{\partial V_1}{\partial \phi_i} =& 
\sum_k n_k\frac{m_k^2}{32\pi^2}
\frac{\partial m_k^2}{\partial\phi_i}\log\frac{m_k^2}{Q^2},\\
\label{eqn:d2V1}
\frac{\partial^2 V_1}{\partial \phi_i \partial\phi_j} =&
\sum_k \frac{n_k}{32\pi^2}\left[
\frac{\partial m_k^2}{\partial\phi_i} \frac{\partial m_k^2}{\partial\phi_j} \left(\log\frac{m_k^2}{Q^2}+1\right) + 
m_k^2\log\left(\frac{m_k^2}{Q^2}\right)\frac{\partial^2 m_k^2}{\partial \phi_i\partial\phi_j} \right].
\end{align}

Plugging eqs.~\eqref{eqn:counter_terms} and \eqref{eqn:d2V1} into \eqref{eqn:F_0_APP}, one finds the contribution of the counter-terms to the effective potential at the electroweak minimum,
\begin{align}
\nonumber
V^{CT}\big|_v=&-\sum_{k}
    \frac{n_{k}}{4\times64\pi^2}
    \Bigg[
        (vI_k)^2
        \left(\log\frac{|m_{k}^2|}{Q^2} + 1\right)
        +
        m_{k}^2
        \log\frac{|m_{k}^2|}{Q^2}
        \left(v^2J_k
        -5vI_k
        \right)
    \Bigg],
\\
&\text{with }
I_k\equiv \frac{\partial m^2_k}{\partial h_1}\Bigg|_v
\text{ and }
J_k\equiv \frac{\partial^2 m^2_k}{\partial h_1^2}\Bigg|_v.
\end{align}

Finally, putting everything together back into eq.~(\ref{eqn:F_0_APP}), including the explicit contributions to $\Delta V_1$, we find
\be\begin{split}
	\label{F_0_APP}
	\mathcal{F}_0 = &~\mathcal{F}_0^{\rm SM} 
				- \frac{v^2}{8}c^2_{\beta-\alpha}\,(m_{H_0}^2 - m_h^2)
				- \frac{m_h^4}{64\pi^2}\left( 3 + \log 2\right) 
				 -\sum_{k} 
	\frac{m_{k,0}^4}{64\pi^2}\left(\log\frac{|m_{k,0}^2|}{Q^2}-\frac{1}{2}\right) \\
	&+ \frac{1}{4\times 64\pi^2}\sum_k\left\{
			(v\,I_k)^2 - 2\,m_k^4 
			+ \left[
				\left( v\,I_k - 2\,m_k^2 \right)^2
				+ m_k^2\,\left(v^2 J_k - v I_k \right)
			\right]\log\frac{m_k^2}{Q^2}
			\right\},
\end{split}
\ee
where the SM vacuum energy of eq.~\eqref{F_0_SM} has been reintroduced and the contribution to the vacuum energy from loops of the SM Higgs and Goldstones, which also occur in $\Delta V_1$, are explicitly subtracted to avoid double counting these terms. Here, $m_{k,0}^2$ denotes a field dependent mass squared evaluated at the origin. This defines the vacuum energy difference of eq.~\eqref{F_0_Final}. 

What remains is to compute the derivatives of the field dependent masses with respect to $h_1$ via the general relations~\cite{Magnus:1985}
\begin{align}\label{eqn:dM}
\frac{\partial m_k^2}{\partial\phi_i} &=\left( \bar{R}\,\frac{\partial M}{\partial \phi_i}\bar{R}^T \right)_{kk}\, ,\\\nonumber
	\frac{\partial^2 m_k^2}{\partial \phi_i\partial\phi_j} &= 	
		\left(\bar{R}\, 
			\frac{\partial^2 M}{\partial \phi_i\partial\phi_j} 
			\bar{R}^T 
		\right)_{kk} 
		+ 2\,\left( 
			\bar{R} 
			\frac{\partial M}{\partial \phi_i}
			\bar{R}^T 
		\right)_{ki}
		\left(m_k^2\,\mathbb{I} - M_{\rm diag}\right)_{ii}^+
		\left(
			\bar{R} 
			\frac{\partial M}{\partial \phi_j}
			\bar{R}^T 
		\right)_{ik}\, ,
\end{align}
where $\bar{R}$ is the orthogonal transformation that diagonalises the scalar mass matrix and $(m_k^2-M_{\rm diag})^+$ denotes the Moore-Penrose pseudoinverse of the diagonal matrix in parenthesis. For such a diagonal matrix, the entries of the pseudoinverse are
\be
	(m_k^2-M_{\rm diag})_{ii}^+ = 
		\left\{ \begin{array}{cc}
			0, & (M_{\rm diag})_{ii} = m_k^2,\\
			\left[ m_k^2-(M_{\rm diag})_{ii} \right]^{-1}, & {\rm else.}
		\end{array}\right.
\ee
Note from eq.~(\ref{eqn:d2V1}) that second derivatives of Goldstone masses always enter multiplied by the Goldstone masses themselves, 
which vanish at the electroweak minimum. So we will not need to compute them.

Defining the quantities
\begin{align}
    \begin{split}
    \Delta m_0^2 &\equiv (m_{H_0}^2-m_h^2)\,,\\
    \mathcal{A}&\equiv\frac{s_\alpha c_\alpha}{s_\beta c_\beta}
    =(c_{\beta-\alpha}+s_{\beta-\alpha}t_\beta)
     (c_{\beta-\alpha}-s_{\beta-\alpha}t_\beta^{-1}),
     \end{split}
\end{align} 
the required mass derivatives are given by
\begin{align}	
\begin{split}        
\label{eqn:IkJk}
v\,I_{G} &= m_h^2 + \Delta m_0^2\,c_{\beta-\alpha}^2  \quad \quad \quad \quad \quad \quad  (\text{Goldstone Bosons})  \\
v\,I_{H^\pm} &= 2m_{H^\pm}^2 + m_h^2\,c^2_{\beta-\alpha} 
						+ m_{H_0}^2\,s^2_{\beta-\alpha}
		-\left[
		2M^2 - \Delta m_0^2 \,\mathcal{A} 
		\right] \\
v^2\,J_{H^\pm} &= v\,I_{H^\pm} +  2 c^2_{\beta-\alpha}s^2_{\beta-\alpha}\frac{(\Delta m_0^2)^2}{m_{H^\pm}^2} \\
v\,I_{A_0} &= v\,I_{H^\pm} - 2\,m_{H^\pm}^2 + 2\,m_{A_0}^2 \\
v^2\,J_{A_0} &= v\,I_{A_0} + 2c^2_{\beta-\alpha}s^2_{\beta-\alpha}\frac{(\Delta m_0^2)^2 }{m_{A_0}^2} \\
v\,I_{h} &= 3\,m_h^2 - c_{\beta-\alpha}^2\,\left[
		2M^2 - \Delta m_0^2 \,\mathcal{A} 
		\right] \\
v^2\,J_{h}  &= v\,I_{h} - \frac{2\,c_{\beta-\alpha}^2\,s_{\beta-\alpha}^2}{\Delta m_0^2} \left[
	2 M^2 - \Delta m_0^2\,\mathcal{A}
	\right]^2 \\
v\,I_{H_0} &= 3m_{H_0}^2 - s^2_{\beta-\alpha}\,\left[ 
		2M^2 - \Delta m_0^2 \,\mathcal{A} 
		\right] \\
v^2\,J_{H_0}  &= v\,I_{H_0} + \frac{2\,c_{\beta-\alpha}^2\,s_{\beta-\alpha}^2}{\Delta m_0^2}
		\left[2 M^2 - \Delta m_0^2\,\mathcal{A}\right]^2\,.
\end{split}
\end{align}

It is easy to show that eq.~\eqref{F_0_APP} simplifies to eq.~\eqref{F0_alignment} in alignment. Through a laborious computation 
one can also show that the $Q^2$ dependence always cancels out, so that $\mathcal{F}_0$ is indeed renormalization scale independent.

\bibliographystyle{JHEP}

\end{document}